\documentclass[12pt]{iopart}

\begin{document}

\title[Gamma-rays from White Dwarfs in Globular Clusters]
{Gamma-rays from electrons accelerated by rotating magnetized White Dwarfs in Globular Clusters}

\author{W. Bednarek}

\address{Department of Astrophysics, University of \L \'od\'z,
              ul. Pomorska 149/153, 90-236 \L \'od\'z, Poland}
\ead{bednar@astro.phys.uni.lodz.pl}
\begin{abstract}
Recently a substantial part of globular cluster population has been established as GeV gamma-ray sources by the Fermi-LAT telescope. 
We investigate possible production of the high energy gamma-rays by relativistic electrons injected from the population of fast rotating, magnetized White Dwarfs within the globular cluster. These electrons
comptonize the radiation field within the globular cluster. 
We conclude that gamma-rays produced by electrons accelerated by the whole population of White Dwarfs within a specific cluster are on the level of detectability of the future Cherenkov telescope array provided that a few thousand of magnetized White Dwarfs have been created uniformly during the lifetime of the globular cluster. 
\end{abstract}

\maketitle

\section{Introduction}

Globular Clusters (GCs) are huge concentrations of old stars, with the total mass in the range $\sim 10^5-10^6$ M$_\odot$, contained within a spherical volume with a radius of a few parsecs. About $\sim$150 known GCs create a spherical halo around the Galaxy with a typical distance scale of $\sim 10$ kpc (Harris 1996).
Since GCs are very old objects, they contain remnants of evolution of stars with masses $>$0.8M$_\odot$, i.e. millisecond pulsars (MSPs), Cataclysmic Variables (i.e. accreting White Dwarfs) and Low Mass X-ray Binaries. The presence of these objects manifest themselves with the large number of the X-ray sources discovered within GCs (e.g. Verbunt 2001). Due to the large stellar content, the radiation field within the GCs is dominated by the optical photons whose energy density can be even $\sim$10$^3$ times larger than the energy density of the Cosmic Microwave Background  Radiation (MBR). 
 
Several GCs have been recently established as sources of GeV $\gamma$-rays in the observations with the Fermi-LAT telescope (Abdo et al.~2009a, Abdo et al.~2009b, Kong et al.~2010, Tam et al.~2011, see for review Bednarek~2011). The $\gamma$-ray spectra of GCs show features very similar to those of the recently discovered population of the millisecond pulsars (MSPs), i.e. very flat spectra with an exponential cut-off at a few GeV (Abdo et al.~2010).
The connection between MSPs and $\gamma$-ray emission from GCs has been recently strengthened by the observation of $\gamma$-ray pulsations from a single MSP within the GC NGC 6624 (Abdo et al.~2011) and also probably within M 28 (Pellizzoni et al.~2009). 
In fact, the production of $\gamma$-rays within GCs by MSPs has been proposed even before the discovery by the Fermi-LAT (Harding et al.~2005, Venter \& de Jager~2008, Venter et al.~2009). However the characteristic cut-offs in the $\gamma$-ray spectra at a few GeV may not be common in the case of all $\gamma$-ray emitting GCs since some of recently discovered objects show also spectra clearly extending above $\sim$10 GeV (Tam et al.~2011).

The possibility of the high energy $\gamma$-ray emission from GCs, extending up to TeV energies, has been postulated in the model considered by Bednarek \& Sitarek~(2007). This model proposes that relativistic leptons which, either escape from the MSP magnetospheres or are accelerated in the MSP winds or in pulsar wind collision shocks, comptonize the stellar radiation and the Microwave Background Radiation (MBR) to the TeV $\gamma$-ray energy range (see also Venter et al.~2009, Cheng et al.~2010, Abramowski et al.~2011a). Up to now, TeV $\gamma$-ray emission has not been discovered from GCs by the Cherenkov telescopes (Anderhub et al.~2009, Aharonian et al.~2009), accept recent report on the existence of the TeV $\gamma$-ray source very close to the GC Ter 5 (Abramowski et al.~2011b). 

It has been argued that $\gamma$-ray fluxes from GCs correlate with their basic parameters. For example, Abdo et al.~(2009a) (and also Hui et al.~2010) show clear correlation of the $\gamma$-ray luminosity with the stellar encounter rate which determines the number of MSPs (and also Cataclysmic Variables) within the specific GC. Hui et al.~(2010) also reports the positive correlation of the $\gamma$-ray luminosity with the metallicity of the GC and finds tendency of its correlation with the soft photon densities within the GC.

Although the connection of the MSP content within the GCs to their GeV $\gamma$-ray emission seems to be 
very likely, we wonder whether other sources within the GCs can not  contribute to the observed $\gamma$-ray flux. In this paper we investigate the possibility of acceleration of electrons within the inner magnetospheres of non-accreting White Dwarfs (WDs). These electrons might diffuse within the GC producing the high energy $\gamma$-rays as a result of scattering of the stellar radiation and MBR in the IC process.
Note that a large population of WDs is expected within GCs since all stars with masses in the range $\sim$0.8-8 M$_\odot$ should already finish their evolution as WDs. The existence of ejecting class of WDs (WD pulsars) has been proposed in the past as a possible explanation of the pulsed X-ray emission from fast rotating WDs (with a periods of a few seconds) within binary systems (see e.g. 1E 2259+586, Paczy\'nski (1990) and Usov~(1993)), or the WD binary AE Aquarii (De Jager~1994, Ikhsanov 1998). Such massive, strongly magnetized and short period WDs could also originate in merging of the two WDs within compact binary system.

\section{White Dwarfs in globular clusters}

As we have noted above stars with masses
(in the range $\sim$0.8-8M$_\odot$) within the GCs should finish their life as the WDs. In order to estimate the number of WDs in a specific GC we use the initial mass function of stars, $dN/dM = N_{\rm 0}M^{-\beta}$, with $\beta = 2.35$ above 0.5 M$_\odot$ (Salpeter 1955, Kroupa~2001) and $\beta = 1.3$ below 0.5 M$_\odot$ (Kroupa~2001). The normalization factor is determined on, $N_{\rm 0}\approx 7\times 10^4$ $M_\odot^{-1}$, by assuming 
the total mass of stars within the GC equal to $M_{\rm GC} = 3\times 10^5$ M$_\odot$ and the minimum mass of stars on 0.08 M$_\odot$. In such a case, the number of WDs, which have appeared within such GC, is estimated on $N_{\rm WD}\sim 7\times 10^4$.    
We conclude that the number of the WDs within a specific GC can be even a few hundred times larger than the number of MSPs within the GC estimated in a similar way. A part of these WDs, which is within the compact binary systems, can merge during the lifetime of the GC. As a result, a population of massive WDs (with the masses close to the Chandrasekhar limit) is expected within the GCs. These WDs should have small radii ($\sim 10^8$ cm), short periods (due to the large angular momentum of the WD-WD binary systems), and strong surface magnetic fields ($\sim 10^9$ G).  

\begin{figure*}
\vskip 7.truecm
\includegraphics{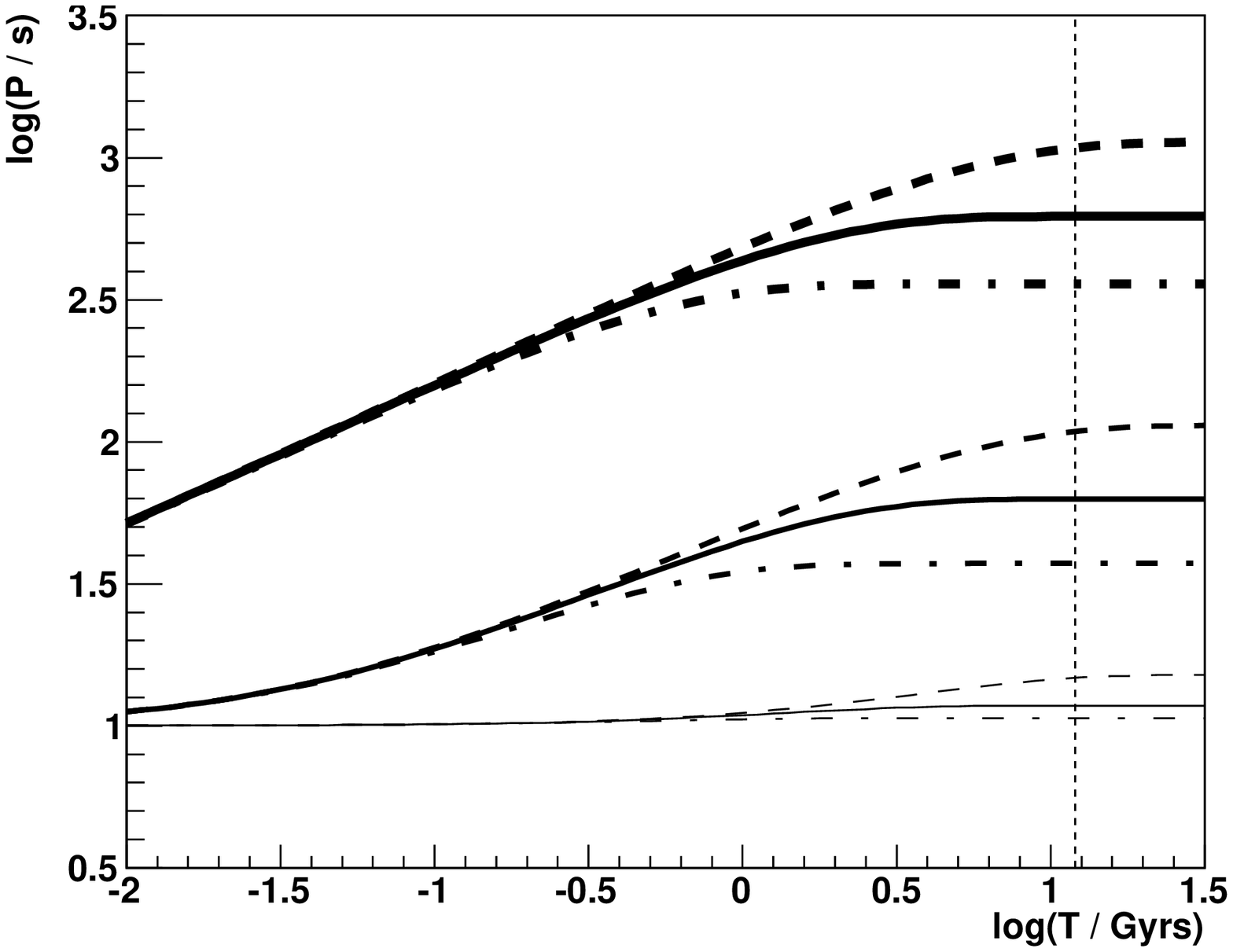}
\includegraphics{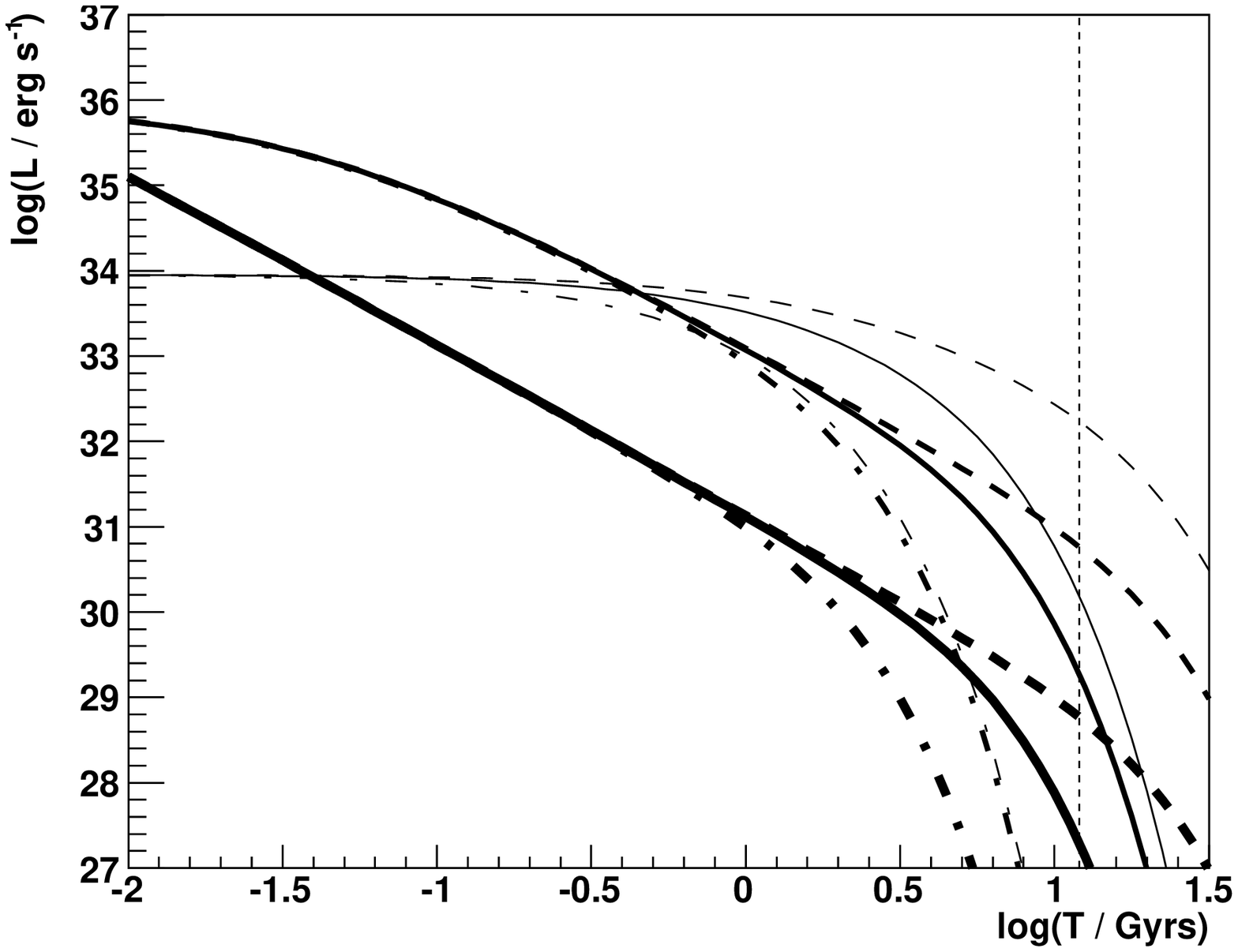}
\includegraphics{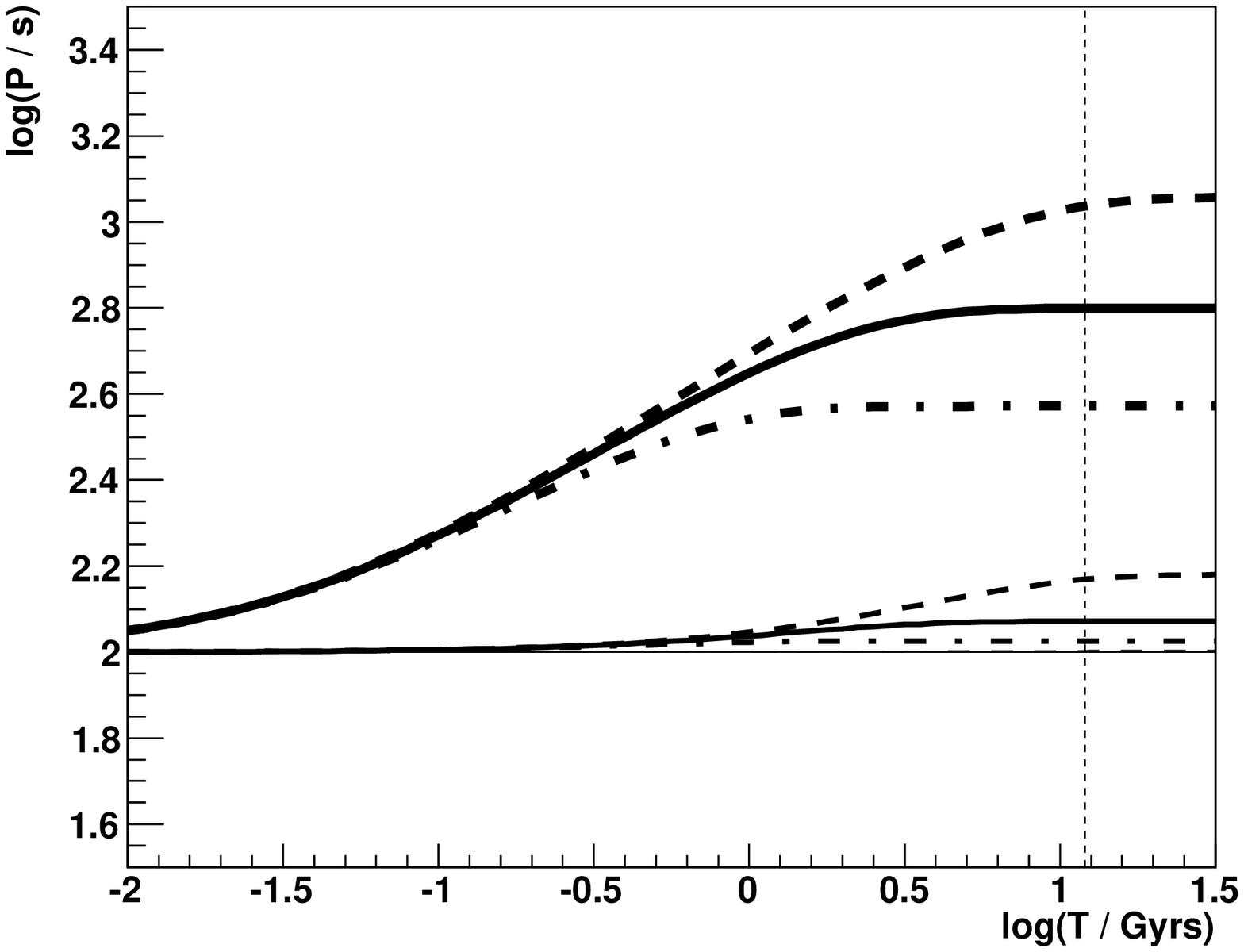}
\includegraphics{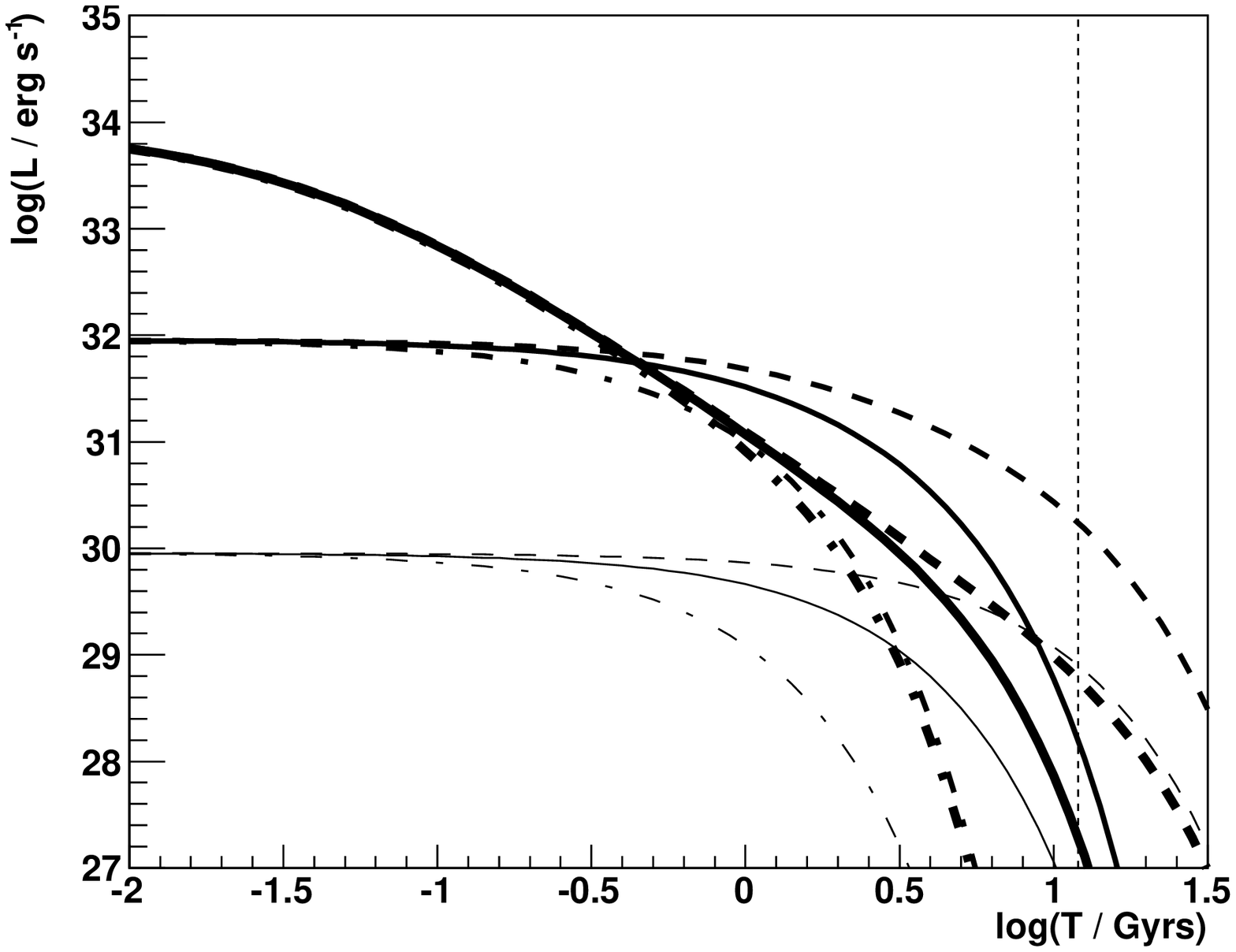}
\includegraphics{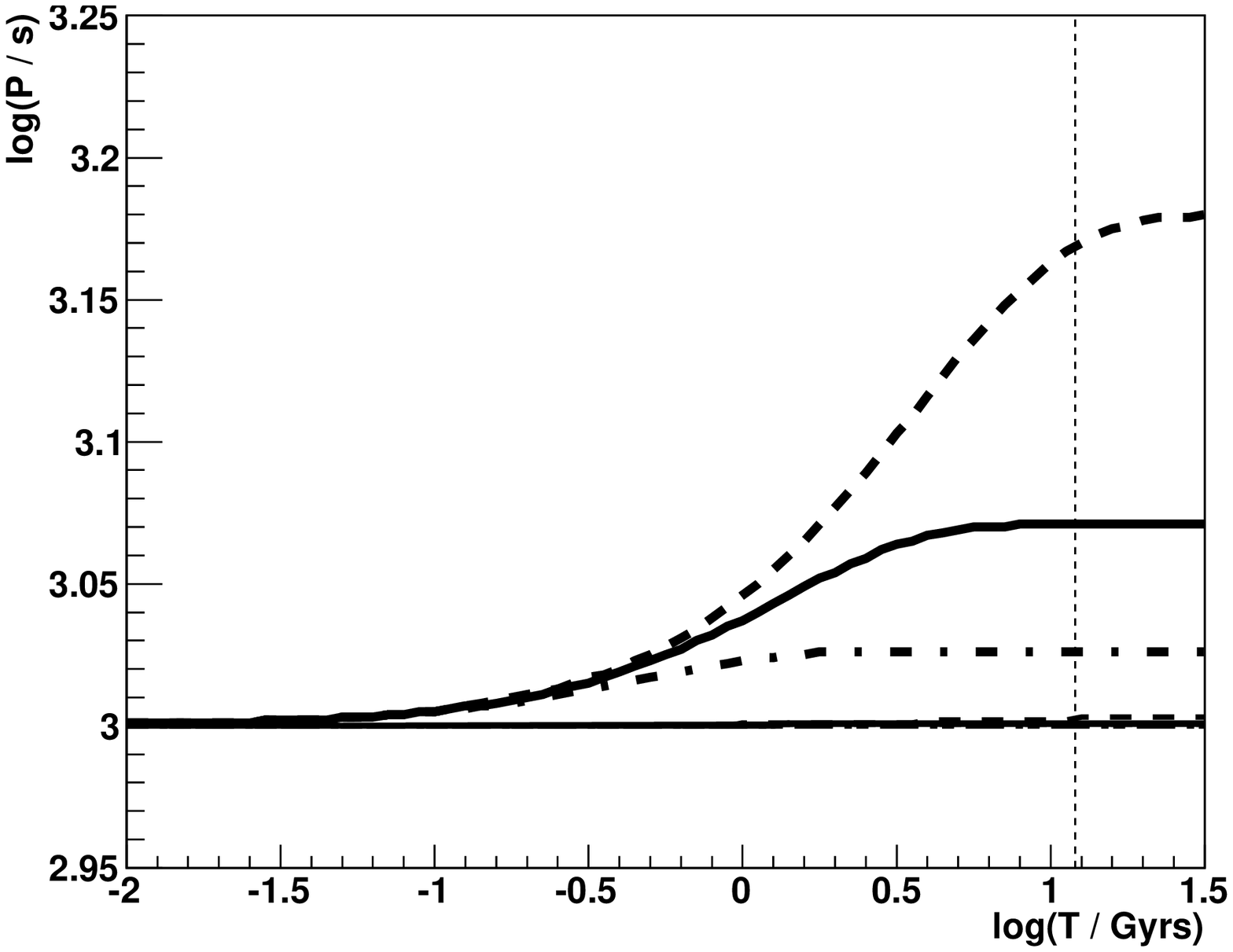}
\includegraphics{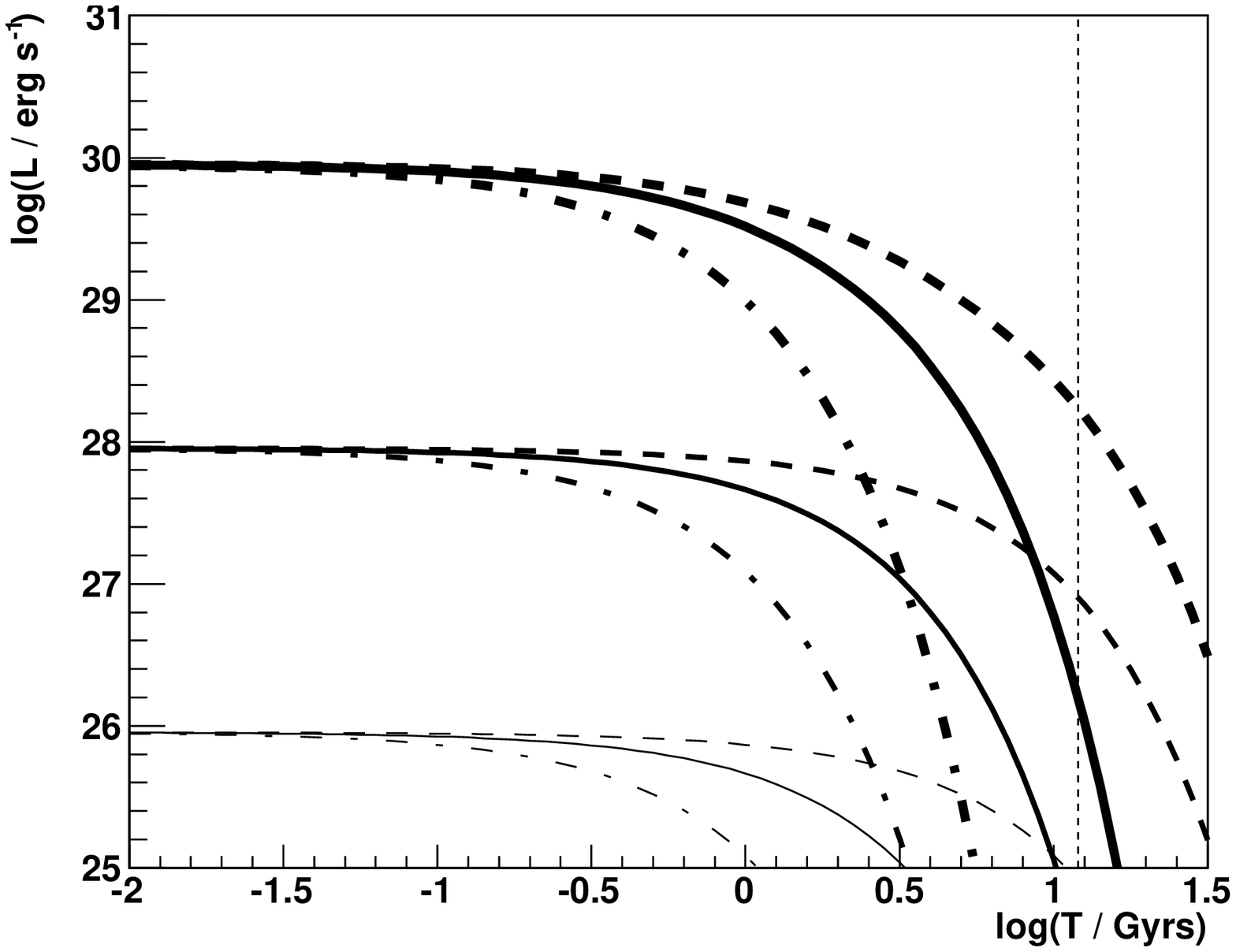}
\caption{The evolution of the rotational period (upper figures) and the rotational energy loss rate (bottom) of the White Dwarf which has been created within the GC soon after its formation and suffered rotational energy losses in a mechanism similar to that observed in rotating neutron stars. The WD has been born with the initial period: $P_{\rm 0} = 10$ s (left figures), 100 s (middle figures), and $10^3$ s (right figures). The WDs with different surface magnetic field strengths are considered: $B_{\rm 0} = 10^7$ G (thin curves), $10^8$ G (middle curves), and $10^9$ G (thick curves). It is assumed that the surface magnetic field of the WD decays on the characteristic time scale: $\tau_{\rm d} = 1$ Gyr (dot-dashed), $3$ Gyrs (solid), and $10$ Gyrs (dashed). The dotted vertical line mark the age of the globular cluster assumed to be equal to 12 Gyrs.}
\label{fig1}
\end{figure*}

In the present paper we are mainly interested in the non-accreting  magnetized WDs (MWDs) since such objects should behave similar to a rotation powered neutron stars (pulsars). Further on we scale the surface magnetic fields of the WDs by $B_{\rm WD} = 10^8B_8$ G and their rotational periods, $P_{\rm WD} = 100P_2$ s since these values seem to be typical for the MWD population. 
We apply the following basic parameters of the White Dwarfs which appeared as a final products of the stellar evolution: their masses 
$M_{\rm WD} = 0.8M_\odot$ and the radii $R_{\rm WD} = 5\times 10^8$ cm.

\subsection{Evolution of the parameters of the WD population}

The rotational energy loss rate of the WD is,
\begin{eqnarray}
\dot{E_{\rm rot}} = {{B_{\rm WD}^2\Omega_{\rm WD}^4R_{\rm WD}^6}\over{6c^3}}\approx 1.5\times 10^{31}B_8^2/P_2^4~~~{\rm erg~s^{-1}},
\label{eq3}
\end{eqnarray}
\noindent
where  the angular velocity of the WD is $\Omega_{\rm WD} = 2\pi/P_{\rm WD}$. 

The total rotational energy badget of the WD is,
\begin{eqnarray}
E_{\rm tot} = 0.5I\Omega_{\rm WD}^2\approx 4\times 10^{47}/P_2^2~~~{\rm erg},
\label{eq6}
\end{eqnarray}
\noindent
where the moment of inertia of the WD is $I = 0.4M_{\rm WD}R_{\rm WD}^2\approx 2\times 10^{50}$ g cm$^{2}$. 
Therefore, for typical parameters, the WD is able to inject relativistic electrons continuously during the characteristic time, 
\begin{eqnarray}
\tau_{\rm WD} = E_{\rm tot}/{\dot E}_{\rm rot}\approx 1.6P_2^2/B_8^2~~~{\rm Gyrs}.
\label{eq7}
\end{eqnarray}
The magnetic WDs are expected to originate from the specific class of Ap type stars which have the masses of the order of $\sim 2$ M$_\odot$. These stars finish their evolution after about $\sim$1 Gyr. Therefore, we expect that at the present age these magnetic WDs are at least $\tau_{\rm MWD}\sim 10$ Gyrs old. 
By comparing this time scale with the characteristic rotational energy loss time scale of the WD with specific parameters ($\tau_{\rm WD} = \tau_{\rm MWD}$), we conclude that these old WDs should have at present the periods longer than,
\begin{eqnarray}
P_2\approx 2.5B_8\tau_{10},
\label{eq8}
\end{eqnarray}
\noindent
where $\tau_{\rm MWD} = 10\tau_{10}$ Gyrs. 
Therefore, WDs, originated soon after formation of the GC, are not able to inject electrons with the power greater than, 
\begin{eqnarray}
\dot{E_{\rm rot}}\approx 10^{28}B_8^{-2}~~~{\rm erg~s^{-1}}.
\label{eq9}
\end{eqnarray}
\noindent
Note however, that the power injected from thousands of such MWDs
within specific GC may become essential.

The estimates made above base on the assumption that the magnetic field of the WDs do not decay during the lifetime of the GCs. This may not be the case. Let us assume that the characteristic decay time of the WD surface magnetic field, $\tau_{\rm dec} = 1\tau_{\rm d}$ Gyrs, can be shorter than the age of the GC. We apply the simple exponential decay law for the surface magnetic field of the WD of the form, 
$B_{\rm WD}(t) = B_{\rm o} \exp(-t/\tau_{\rm d})$, where $B_{\rm o}$ is the surface magnetic field of the WD at birth. In such a case, the period of the WD, at the time, t, after its birth, can be obtained from the comparison of the differentiate of the rotational energy losses of the WD (Eq.~2) with the energy loss rate of the rotating magnetic dipole (Eq.~1). It is given by,
\begin{eqnarray}
P^2(t) = P^2_{\rm 0} + 3\times 10^3B_8^2\tau_{\rm d}
[1 - \exp(-2t/\tau_{\rm d})]~~~{\rm s^2},
\label{eq10}
\end{eqnarray}
\noindent
where the initial period of the WD is marked by $P_{\rm 0}$.
We show the example evolutionary tracks of the basic parameters of the WDs as a function of time, i.e. their periods and rotational energy loss rates, 
for a few different initial periods, and characteristic decay times of their surface magnetic field (see Fig.~1).
Note, that the WDs, which have been born with short periods and strong magnetic field surfaces, may have at present the energy loss rates even lower than those born with longer periods and weaker surface magnetic fields provided that their decay time scale for the surface magnetic field differ significantly.

We have also considered the magnetized WDs which could be created as a result of WD-WD mergers. As noted above these WDs should have larger masses ($\sim 1.4$ M$_\odot$) and smaller radii ($\sim 10^8$ cm). 
The rotational energy loss rates of this population of WDs can be estimated on,
$E_{\rm rot}\approx 5.8\times 10^{27}B_8^2/P_2^4~~~{\rm erg~s^{-1}}$. The evolution of their rotational periods is then determined by 
$P^2(t) = P^2_{\rm 0} + 4.6B_8^2\tau_{\rm d}
[1 - \exp(-2t/\tau_{\rm d})]~~~{\rm s^2}$.
Note that WDs, created in  merger events, lose energy at a lower rate (for this same initial surface magnetic fields and periods) resulting in a slower evolution of their periods in time.
This effect is due to the strong dependence of the rotational energy loss rate on the radius of the WD (see Eq.~1).

\subsection{Initial parameters of the WD population}

WDs have to be very common objects in GCs since most of the stars with the masses $\sim$M$_\odot$ have already finished their evolution during the period of several Gyrs. In fact, many faint X-ray sources have been discovered in GCs by the X-ray satellites. Many of these X-ray sources are identified with accreting WDs within compact binary systems, so called Cataclysmic Variables (e.g. Grindlay et al.~2001, Heinke~2011). However, far more WDs within GCs should be isolated since the fraction of stars within the binary systems is estimated to be typically as law as $\sim$10$\%$ (Ivanova et al.~2005).

The initial parameters of the population of the WDs within the GCs are unknown. The stellar evolution models conclude that all stars with masses in the range between $0.8-8$ M$_\odot$ should end their life as the WDs. 
So then, their total number in specific GC can be estimated from the knowledge on its mass. We may only have some insights into the parameter range of WDs, their periods and surface magnetic fields, from the studies of the WDs in the surrounding of the Sun. The observations show that magnetic WDs comprise $\sim$5$\%$ of all WDs while magnetic CVs comprise $\sim$25$\%$ of all CVs (Wickramasinghe \& Ferrario~2000). The average surface magnetic field
is $<$log B(MG)$> = 1.193\pm 0.741$ for the isolated WDs (see Fig.~23 in Wickramasinghe \& Ferrario~2000). On the other hand, the magnetic field strength of the sub-class of accreting WDs in binary systems called AM Her lays in the range $10-60$ MG with the mean $38\pm 6$ MG. The lack of WDs with stronger field may be the selection effect and the WDs with weaker field belong to other sub-class of accreting WDs, so called Intermediate Polars (Wickramasinghe \& Ferrario~2000). The periods of the observed isolated MWDs peaks at $10^4$ s having broad distribution down up to $\sim 10$ s. 
These observations of the local population of WDs may give some insight into the parameters of the WDs but may also not correspond to the reality due to the different evolutionary tracks of the WDs in dense cores of GCs.

The initial periods of the WDs which appear as a result of the evolution of Ap and Bp type stars can be estimated assuming that the angular momentum of the star is more or less conserved. For example, let us consider Ap type star which has radius $\sim 10^{11}$ cm and the rotational velocity $\sim 100$ km s$^{-1}$. If the WDs form with the radius $10^8$ cm, then their initial periods might be as short as a part of a second. For less massive stars (similar to our Sun: the radius $R_\odot = 7\times 10^{10}$ cm and the rotational velocity $V_\odot = 2$ km s$^{-1}$, the created WD should have the initial period of the order of $\sim 100$ s (assuming conservation of the angular momentum). It seems likely that created WDs can rotate with the initial periods in the range $\sim 1-100$ s.

The rotational period of the WD created during the merger event of two WDs can be estimated by assuming that the angular momentum of the WD binary system has been conserved during the merger. Let us assume that the initial period of the WD-WD binary system was $\tau_{WD} = 1\tau_{\rm d}$ days and the masses of the WDs were $M_{\rm WD} = 0.7M_\odot$. Then, the rotational period of such a WD from merger can have the period as short as, 
\begin{eqnarray}
P_{\rm WD} = {{2\pi R_{\rm WD}}\over{2.5^{1/2}}}\left[{{32\pi^2}\over{G^2M_{\rm WD}^2\tau_{\rm WD}}}\right]^{1/6}\approx {{0.3 R_8}\over{M_{\rm WD}^{1/3}\tau_{\rm d}^{1/6}}}~~{\rm s}, 
\label{eq11}
\end{eqnarray}
\noindent
where the radius of the WD is $R_{\rm WD} = 10^8R_8$ cm.   
It is likely that the WDs rotating with the periods of the order of seconds can be created in the merger events of WDs in binary systems.

The strong magnetic fields of the WDs in GCs might also decay during their long lifetime. The calculations show that the characteristic decay time depends on the WD mass. It is in the range $\sim$1-10 Gyrs (Wendell et al.~1987, Muslinov et al.~1995). Initially much weaker higher order poloidal component of the magnetic field may decay slower in respect to the dipolar component reaching the characteristic decay time scale of $\sim$10 Gyrs.

\section{Relativistic electrons from White Dwarfs}

We consider two populations of MWDs. The first one appears within the GCs as a final product of the evolution of stars with masses in the range $0.8-8$M$_\odot$. The second one is due to the mergers of two WDs within the compact binary system. These two groups of WDs have different masses and radii. They differ significantly in the moment of inertia.
This results in a different evolutionary tracks of their periods in time.

\subsection{WDs from stellar evolution}

The maximum Lorentz factors of electrons accelerated in the WD inner magnetosphere along the electric field lines induced through the open magnetic field can be estimated from (Goldreich \& Julian~1969),
\begin{eqnarray}
\gamma_{\rm max} = \xi eB_{\rm WD}\Omega_{\rm WD}^2R_{\rm WD}^3/(2m_ec^4)\approx 1.6\times 10^7\xi B_8P_2^{-2}. 
\label{eq12}
\end{eqnarray}
\noindent
where $\xi$ is the acceleration efficiency, $e$ is the electron charge, $c$ is the velocity of light, and $m_{\rm e}$ is the electron rest mass. In fact, in a realistic case the energy of accelerated particles grows gradually from the surface of the WD up to some maximum value, e.g. as obtained in a space charge limitted outflow model considered by Arons \& Scharlemann~(1979) or Harding \& Muslimov~(1998). Moreover, in this model the acceleration process along specific magnetic field lines is expected to occur differently. In the present paper we do not consider the details of a specific acceleration scenario since the main topic of our considerations is the propagation of electrons and their radiation processes already within the GC. Therefore, we simply scale the maximum possible electron Lorentz factors from specific WDs with the acceleration efficiency parameter which is assumed much less than unity.

\begin{figure*}
\vskip 5.5truecm
\includegraphics{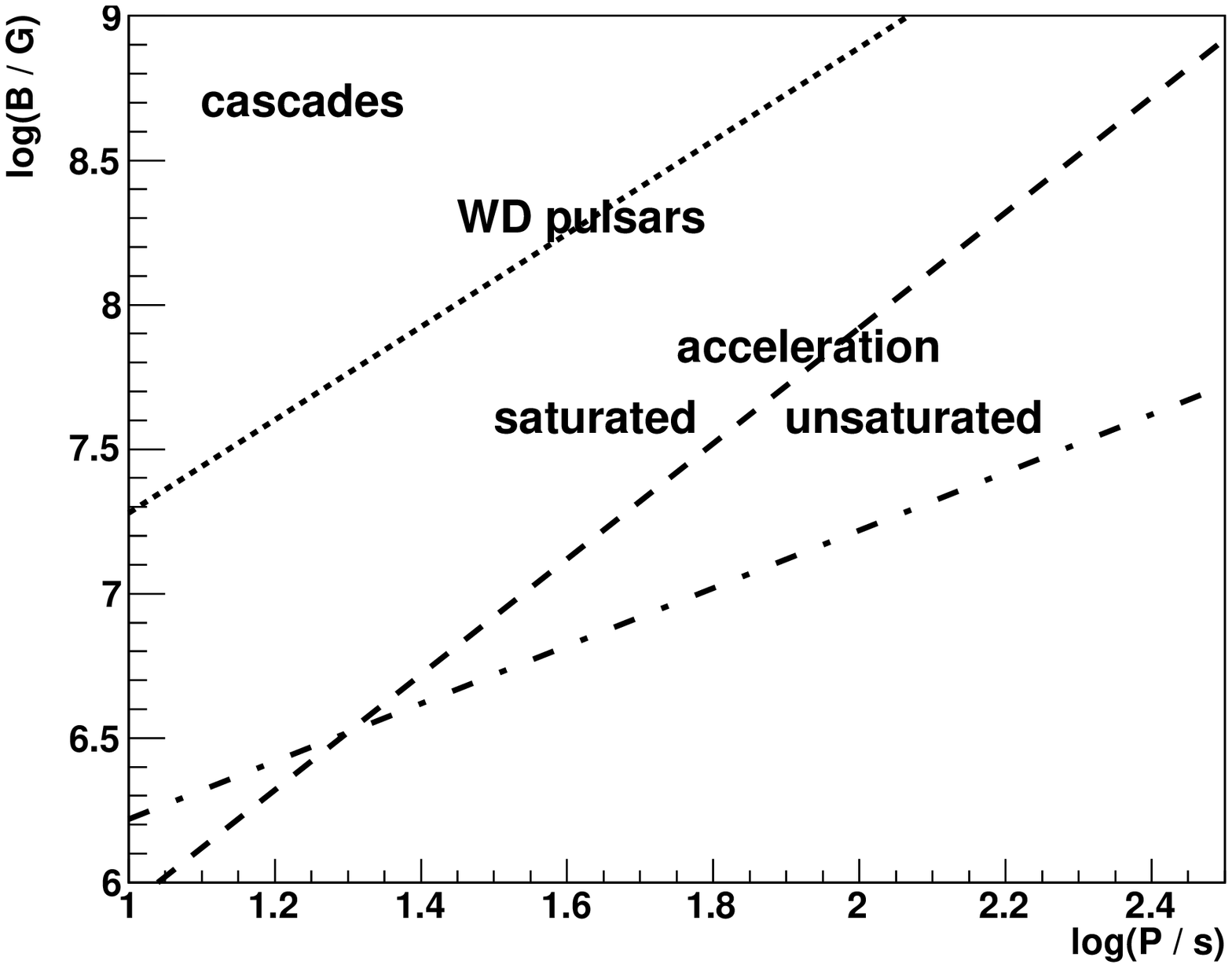}
\includegraphics{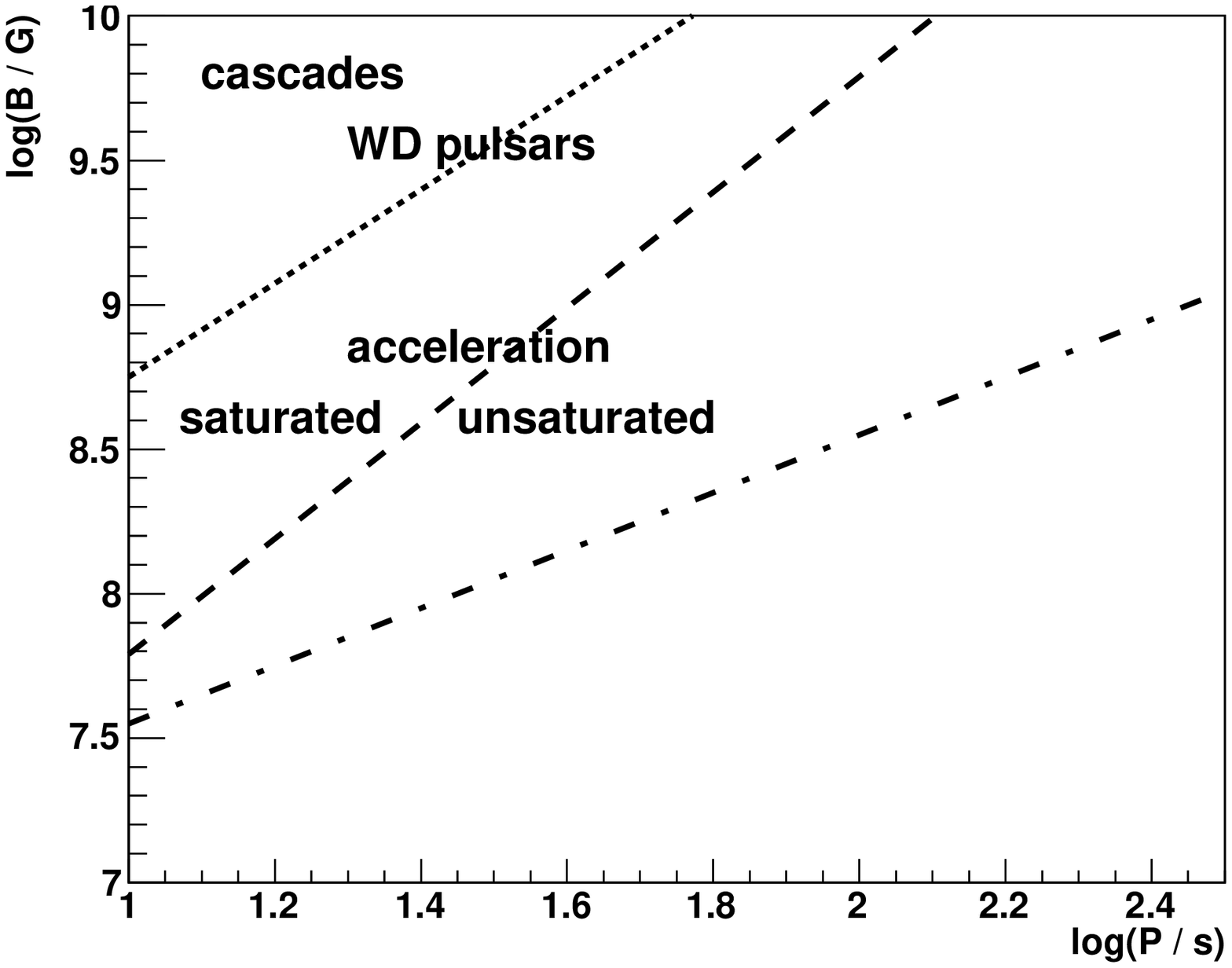}
\caption{Conditions for acceleration of electrons in the inner White Dwarf magnetosphere as a function of the surface magnetic field of the WD, $B$, and its rotational period, $P$: for the WDs created as a result of stellar evolution (figure on the left) and originated from the WD-WD mergers within the compact binary systems (figure on the right). The condition for $e^\pm$ pair production by the curvature $\gamma$-rays in the magnetic field of the inner WD magnetosphere (dotted line). The condition for the saturation (or not saturation) of the electron acceleration by the curvature energy losses (dashed line). The WD should operate as a pulsar (due to the conversion of electron energy to radiation already in the inner magnetosphere)
for parameters above the dashed line. The final parameters of the WD (see Eq.~\ref{eq8}) which has been created within the Globular Cluster and slowed down during its lifetime (12 Gyrs) due to the rotational energy losses are marked by the dot-dashed line.}
\label{fig2}
\end{figure*}

The acceleration process of these electrons can be limited by their energy losses on the curvature radiation due to the curvature of the magnetic field lines in the inner WD magnetosphere. We estimate the parameter range of the WDs for which the curvature energy losses are not able to limit their acceleration. The following condition has to be fulfilled,
$\lambda_{\rm cr} > R_{\rm LC}$, where $\lambda_{\rm cr} = m_{\rm e}c^2\gamma_{\rm max}/{\dot E}_{\rm cr}$ is the electron energy loss mean free path on the curvature radiation. The curvature energy losses are given by,
\begin{eqnarray}
{\dot E}_{\rm cr} = {{2e^2\gamma_{\rm max}^4}\over{3R_{\rm cr}^2}}.
\label{eq13}
\end{eqnarray}
\noindent
The average curvature radius of the magnetic field lines is estimated from $R_{\rm cr} = \sqrt{R_{\rm WD}R_{\rm LC}}\approx 1.5\times 10^{10}P_2^{1/2}$ cm, where $R_{\rm LC} = cP_{\rm WD}/2\pi$ is the light cylinder radius. This approximation  is good for the magnetic field line with the purely dipolar structure attached to the outer rim of the polar cap and not far away from the WD surface. The curvature radii of the magnetic lines closer to the magnetic pole are expected to be larger. Therefore, curvature energy losses along those lines are weaker, allowing the acceleration of electrons closer to $\gamma_{\rm max}$.

The curvature loses are not able to limit the electron acceleration in the WD magnetosphere (i.e. $\lambda_{\rm cr} > R_{\rm LC}$) for the range of parameters fulfilling the condition,
\begin{eqnarray}
B_8 < 0.84P_2^{2}/\xi. 
\label{eq14}
\end{eqnarray}
\noindent
If Eq.~\ref{eq14} is not fulfilled, then electrons are injected from the inner WD magnetosphere (through the light cylinder radius) with the Lorentz factors, $\gamma_{\rm sat}$, which can be clearly lower than $\gamma_{\rm max}$.
This limitting Lorentz factor can be estimated from the condition $\lambda_{\rm cr}(\gamma_{\rm sat})\approx R_{\rm LC}$. It is equal to,
\begin{eqnarray}
\gamma_{\rm sat}\approx 1.4\times 10^7.
\label{eq15}
\end{eqnarray}
\noindent
Note that it is independent on the WD parameters in the case of the dipole model for the magnetic field structure in the inner magnetosphere of the WD.
Electrons with such Lorentz factors can produce curvature photons with the characteristic energies,
$E_\gamma^{\rm cr} = 3ch \gamma_{\rm sat}^3/(4\pi R_{\rm cr})\approx 5.3P_2^{-1/2}$ MeV (where $h$ is the Planck constant), i.e. the curvature photons should have characteristic energies in the MeV range.
Since the acceleration of electrons can occur along different magnetic field lines, electrons are expected to be injected from the WD magnetospheres with the Lorentz factors in the range
$\gamma_{\rm sat}$ and $\gamma_{\rm max}$. The exact process of acceleration of particles in the magnetospheres of rotating magnetized objects is not to the end known and its consideration is not the main topic of this paper. We concentrate on the propagation and radiation processes of electrons within GCs with the lower value of the Lorentz factors given by Eq.~\ref{eq12} and~\ref{eq15}.

The curvature $\gamma$-rays can be efficiently converted into $e^\pm$ pairs in the magnetic field of the WD if the following condition is fulfilled (Ruderman \& Sutherland~1975),
\begin{eqnarray}
{{E_\gamma^{\rm cr}}\over{2m_{\rm e}c^2}}{{B}\over{B_{\rm cr}}} > {{1}\over{15}},
\label{eq16}
\end{eqnarray}
\noindent
where $B_{\rm cr} = 4.4\times 10^{13}$ G is the critical magnetic field strength.
Electrons with the Lorentz factors, $\gamma_{\rm sat}$ following the magnetic field with curvature $R_{\rm cr}$, can not fulfil the above condition for realistic values of the surface magnetic field of the WD since it requires
$B_8 > 5.4\times 10^3P_2^{1/2}$. Even if the magnetic field is dominated by the higher order components, having the curvature radius comparable to the radius of the WD, the absorption of the curvature photons in the WD magnetic field can only occur for  $B_8 > 180$.
This condition is not fulfilled by considered by us WDs.

Only in the case of almost instantaneous acceleration of electrons, relatively close to the WD surface (without curvature losses important),
the energies of curvature $\gamma$-rays produced by these electrons can be as large as,
\begin{eqnarray}
E_\gamma^{\rm cr} = {{3}\over{4\pi}}{{ch}\over{R_{\rm cr}}}\gamma_{\rm max}^3\approx 8\xi^3B_8^3P_2^{-13/2}~~~{\rm MeV}.
\label{eq17}
\end{eqnarray}
\noindent
Applying the curvature radius of the WD magnetic field lines in the inner magnetosphere as expected in the dipole model (see above), we get the limit, 
\begin{eqnarray}
B_8 > 7.8\xi^{-3/4}P_2^{13/8}, 
\label{eq18}
\end{eqnarray}
\noindent
for which curvature $\gamma$-rays can create $e^\pm$ pairs in the WD magnetic field. As a result, $e^\pm$ pair cascades in the magnetic field can develop along certain magnetic field lines within the WD inner magnetosphere. These pair production process can quench farther acceleration of electrons to large energies along these lines due to the saturation of the electric field. 
Note that for large range of the WD parameters (B, P), even in the case of almost instantaneous acceleration of electrons, the electric field can not be saturated by the $e^\pm$ pair production in the inner magnetosphere which appears as a result of the conversion of curvature $\gamma$-rays to $e^\pm$ pairs (see Fig.~2). 
We conclude that the WDs with the parameters in the range limitted by the dotted and dashed lines in Fig.~2 (the case of saturated acceleration) should manifest themselves as the MeV pulsars (see also Paczy\'nski 1990, Usov~1993).
We can distinguish a few regimes for acceleration of electrons in the 
inner WD magnetosphere. For the parameters above the dotted line, the energies of curvature $\gamma$-rays might be sufficient enough for their convertion into $e^\pm$ pairs in the magnetic field of WD. Produced $e^\pm$ pairs can saturate the electric field induced in rotating magnetosphere preventing farther acceleration of $e^\pm$ pairs to energies comparable to the maximum potential drop through the open magnetosphere. As a result, the primary electrons are accelerated to relatively low energies
but plenty of $e^\pm$ pairs escape through the light cylinder.
Below the dotted line, electrons can be accelerated to large energies. Above the dashed line, their acceleration is limited by the curvature energy losses (so-called saturated acceleration). Below the dashed line, electrons reach maximum energies corresponding to the full potential drop available along the open field lines. For the parameters of the WDs below the dotted line, electrons are injected through the light cylinder with large Lorentz factors. The rotating WD produce collimated pulses of radiation due to the curvature radiation, which for the parameters below the dotted line, are expected up to the soft $\gamma$-ray energies. The WDs with parameters above the dotted line might produce plenty of $e^\pm$ pairs in the inner magnetosphere which are expected to emit coherent synchrotron radio emission. Thus, these WDs might resemble classical radio pulsars. 
Note however, that this can happen only in the case of very efficient acceleration of electrons close to the WD surface without significant
energy losses.

The electrons accelerated in the WD magnetosphere takes a part, $\eta$, of the rotational energy lost by the WD (described by Eq.~1). Then, the power contained in relativistic electrons can be related to the WD energy loss rate by,
$L_{\rm e}^{\rm unsat} = \eta E_{\rm rot}$ in the case of the WDs with unsaturated acceleration of electrons in the inner magnetospheres and by $L_{\rm e}^{\rm sat} = \eta (\gamma_{\rm sat}/\gamma_{\rm max})E_{\rm rot} $ in the case of saturated acceleration. We estimate the number of injected relativistic  electrons by normalizing their energies to the estimated above
power, $L_{\rm e}^{\rm unsat}$ and $L_{\rm e}^{\rm sat}$, respectively.

\subsection{WDs from mergers}

The MWDs created in the merger events have similar magnetic moments to those created from the magnetic stars but their rotational periods are expected to the clearly shorter, due to the large angular momentum of the binary system. Thus, the curvature of the magnetic field lines should be smaller due to their larger masses and smaller radii. Therefore, in principle they could accelerate electrons to larger energies in their inner magnetospheres (see Eq.~\ref{eq12}) but the condition for the saturation of the acceleration should be more restrictive. As the basic parameters of WDs from mergers, we apply the following values: $M_{\rm WD} = 1.4M_\odot$ and $R_{\rm WD} = 10^8$ cm.

The condition for saturation of electron acceleration now becomes
(see also this condition for the WDs from stellar evolution given by Eq.~10),
\begin{eqnarray}
B_8 > 62P_2^2/\xi. 
\label{eq19}
\end{eqnarray}
In the case of saturated acceleration, electrons move in the inner magnetosphere with the equilibrium Lorentz factor, $\gamma_{\rm sat}\approx 8\times 10^6$, and the characteristic energies of curvature $\gamma$-rays are $E_\gamma^{\rm cr}\approx 2$ MeV. The condition for developing cascades in the WD inner magnetosphere (Eq.~\ref{eq16}) becomes the following,
\begin{eqnarray}
B_8 > 240P_2^{13/8}, 
\label{eq20}
\end{eqnarray}
These conditions for the acceleration of electrons in the magnetospheres of MWDs created in the merger events are also shown in Fig.~2. 
The characteristic 'active' time of the WDs within the GC, estimated from the rotational energy loss time, is $\tau_{\rm WD}\approx 130P_2^2/B_8^2$ Gyrs (see for comparison Eq.~\ref{eq7}). We conclude that the WDs from mergers of the binary WDs created soon after the formation of the GC should have at present rotational periods longer than $P_2\approx 0.28B_8$. 
They are relatively slowly rotating objects at the present time (i.e. 12 Gyrs after their formation) even if their surface magnetic field is of the order of $10^9$ G.

\section{Propagation of electrons in globular clusters}

Relativistic electrons, injected by the rotation
powered WDs within the GC, diffuse slowly in the GC magnetic field. 
We apply a simple diffusion model developed for the case of injection of electrons from the MSP population within the GC (see Bednarek \& Sitarek~2007). Assuming the Bohm diffusion, electrons with specific energy, $E_{\rm e}$, stay within the GC for the average time,
\begin{eqnarray}
t_{\rm diff} = R_{\rm h}^2/D_{\rm diff},
\label{eq21}
\end{eqnarray}
where $D_{\rm diff} = R_{\rm L}c/3$ is the Bohm diffusion coefficient, $R_{\rm L} = cp/eB_{\rm GC}\approx 3\times 10^{14}E_{\rm TeV}/B_{\rm -5}$ cm is the Larmor radius of electrons in the GC magnetic field $B_{\rm GC} = 10^{-5}B_{\rm -5}$ G, $p$ and $E_{\rm e} = 1E_{\rm TeV}$ TeV are the electron momentum and energy respectively, and $R_{\rm h}$ is the half mass radius of the GC. For typical parameters of the GCs, it has been shown (Bednarek \& Sitarek~2007) that relativistic electrons spend enough time for frequent collisions with thermal photons. 
The magnetic field within GC is produced by a large number of compact objects (such as stars, WDs, millisecond pulsars). Therefore, the volume of the GC is expected to be feeled with a strong turbulent magnetic field without any dominant ordered component. In such conditions the Bohm limit for the diffusion process seems to give correct description.
In the case of other dependence of the diffusion coefficient on electron energy (e.g of the form
$D_{\rm diff}\propto E^{\rm (1/3\div 2/3)}$), the large energy electrons diffuse slower than considered in the Bohm limit. However, this should not introduce significant effect on the gamma-ray spectra since electrons even with 
maximum considered energies cool efficiently within the volume of the GC (the process is followed up to 10 pc, see Fig.~5 in Bednarek \& Sitarek 2007 for the dependence of the $\gamma$-ray spectra on the diffusion distance).
Note that electrons diffusing outside the GC move through varying radiation field produced by stars. Therefore, in order to calculate the $\gamma$-ray spectra from GCs, we apply the Monte Carlo method which helps us to follow the processes of energy losses of electrons and production of $\gamma$-rays in the IC scattering of stellar radiation field and the MBR which is considered with the use of the full Klein-Nishina cross section. Density of stellar radiation decreases from the center of the GC as shown in Fig.~1 in Bednarek \& Sitarek~(2007).

\section{Gamma-rays from White Dwarfs in globular clusters}

We calculate the $\gamma$-ray spectra produced by relativistic electrons
in the inverse Compton process applying the general formula given by Blumenthal \& Gould (1970, see Eq.~2.48 in this paper). 
In order to have impression about the possible contribution from WDs with different parameters we first calculate the $\gamma$-ray spectra expected from electrons which are injected by a single WD with specific parameters (period, surface magnetic field). Electrons reach energies described by Eq.~(\ref{eq12}) or~(\ref{eq15}). They contain the power given by Eq.~1. We consider the cases with possible evolution of the basic parameters of the WD during the lifetime of the GC, i.e. assuming that the WD appears soon after the formation of the GC, the surface magnetic field decay on a specific time scale and the period of the WD evolves due to the rotational energy losses according to Eq.~\ref{eq10}. 
We apply the model for the stellar radiation field within the GC as expected for its typical parameters (see also Bednarek \& Sitarek~2007). It is assumed that the GC is at the distance of 7 kpc, have the mass $3\times 10^5M_\odot$, within the region with the core radius $R_{\rm c} = 1.6$ pc and the half mass radius $R_{\rm h} = 3$pc. The diffusion process of electrons is followed in the region of 10 pc around the GC. Note that for typical masses of GCs in the range $10^5-10^6$ M$_\odot$, the stellar radiation field within the GC does not have a strong effect on the producted $\gamma$-ray fluxes and spectra in the IC process since electrons cool efficiently already within the GC (see for detailed comparison Fig.~2 in Bednarek \& Sitarek~2007).
The magnetic field within the GC is taken to be equal to $3\times 10^{-6}$ G. The $\gamma$-ray fluxes, expected from such example GC, are obtained as due to the comptonization of the stellar radiation and the MBR by electrons injected by a single WD (see Fig.~3).

\begin{figure*}[t]
\vskip 3.7truecm
\includegraphics{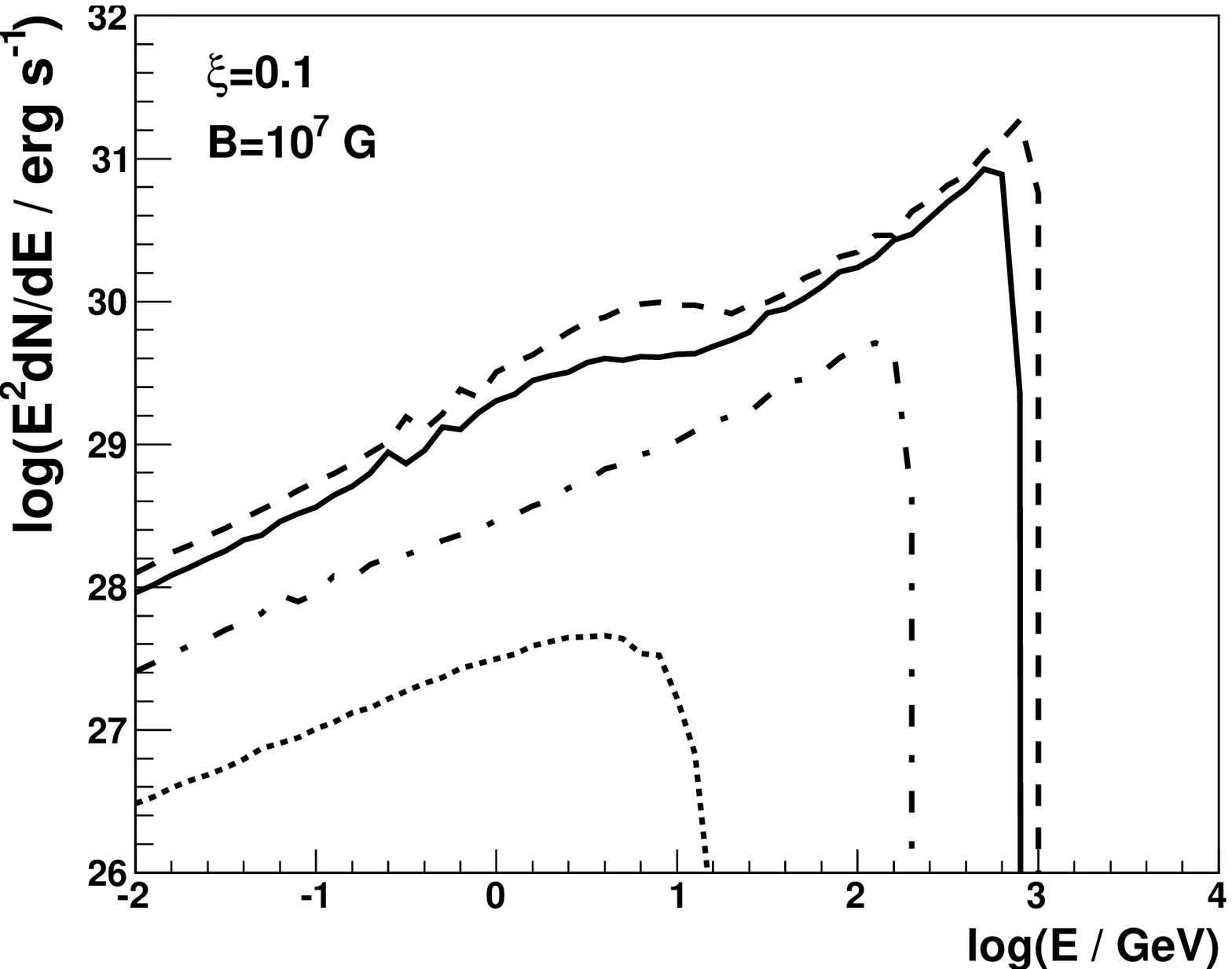}
\includegraphics{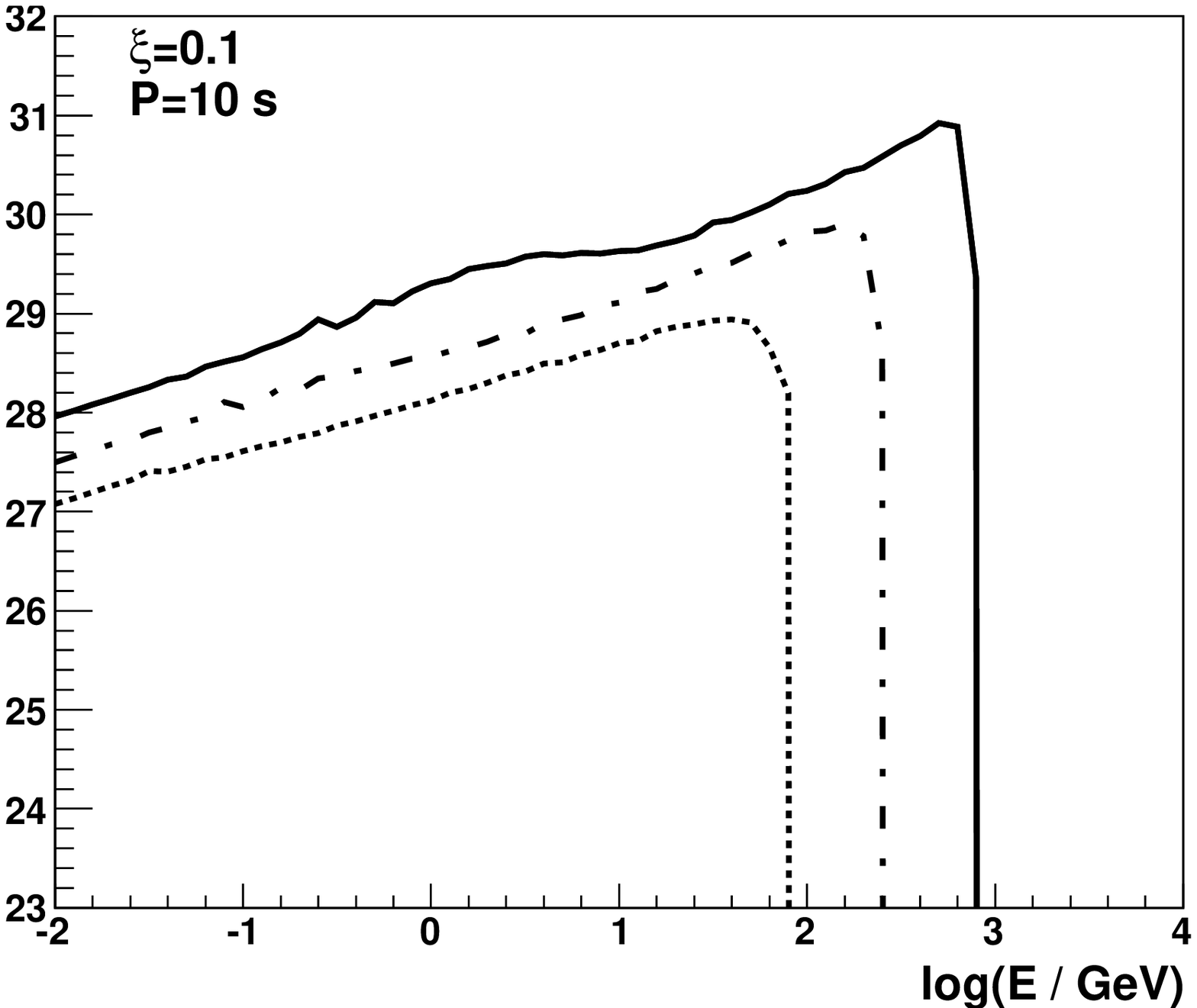}
\includegraphics{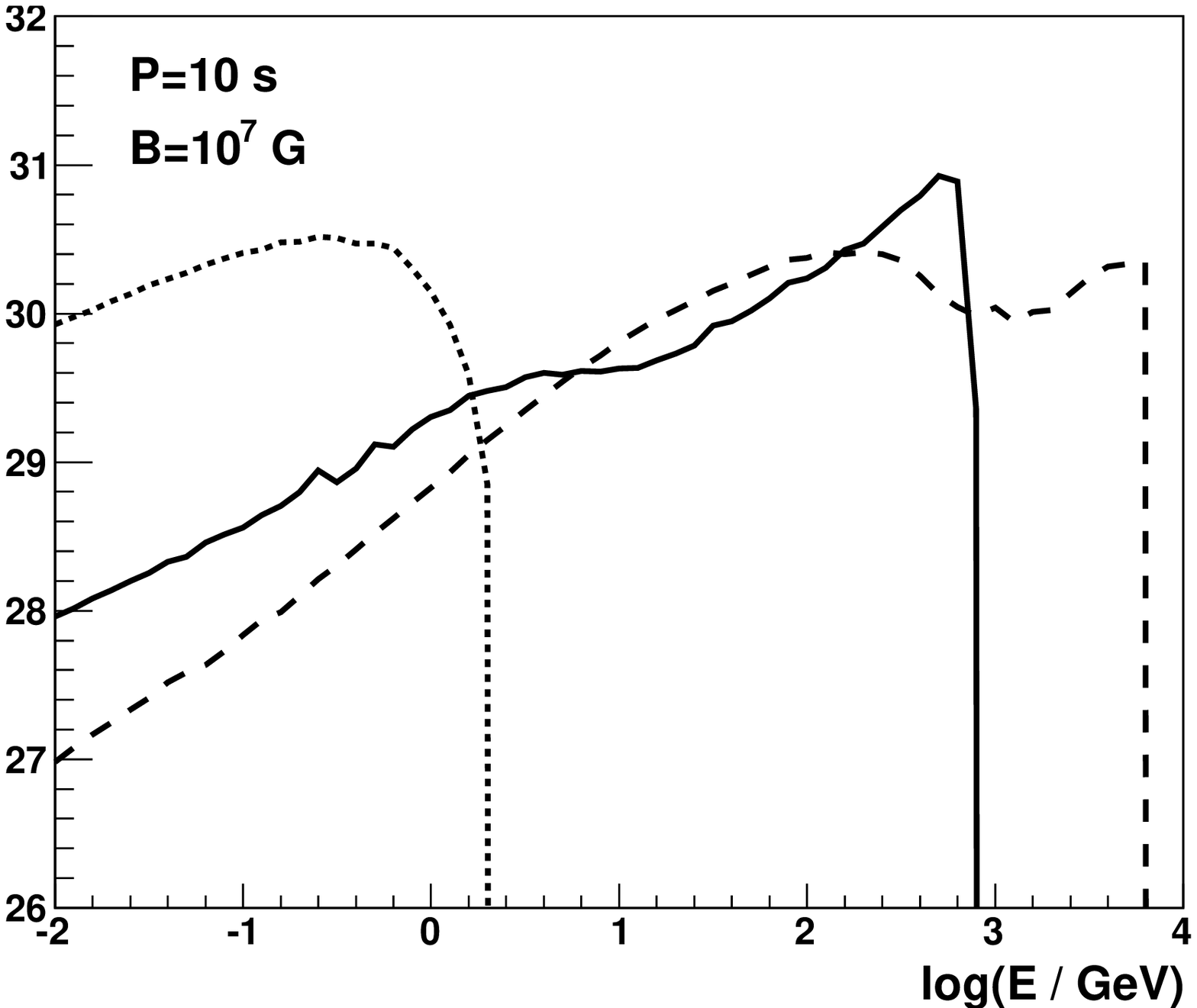}
\caption{Gamma-ray spectra produced by electrons in the IC scattering of 
the stellar radiation and the MBR within the GC. Electrons are injected 
with energies given by Eq.~(8) or~(11) by the WDs. The spectra are shown for different initial period of the WD: $P = 1$ s (dashed curve), $10$ s
(solid), $30$ s (dot-dashed), and $100$ s (dotted). The other parameters are fixed on $B = 10^7$ G, $\xi = 0.1$ (left figure). The spectra for different initial surface magnetic field strengths of the WDs, $B = 10^6$ G (dot-dashed), $10^7$ G (solid), $10^8$ G (dotted), and $P = 10$ s, $\xi = 0.1$ are shown in the middle figure.
The spectra for different acceleration efficiencies, $\xi = 1$ (dashed), 0.1 (solid), $0.01$ (dotted)), and $P = 10$ s and $B = 10^7$ G are shown in the right figure. All  spectra are normalized to the product of the number of WDs created within the GC times the energy conversion efficiency, $N_{\rm WD}\times \eta = 1$. The magnetic field strength within the GC is equal to $B_{\rm GC} = 3\times 10^{-6}$ G.
The surface magnetic field of the WDs is assumed to decay on the time scale of $\tau_{\rm dec} = 10$ Gyrs.}
\label{fig3}
\end{figure*}
\subsection{White Dwarfs from stellar evolution}

As we have discussed above,  WDs created as a final products of the stellar evolution can likely have the initial periods of the order 1-100 s. A significant part of them (about $\sim$10$\%$ of observed WDs) can also have the surface magnetic fields with the strength of the order of $10^7$ G. Therefore, we consider such values for the initial parameters of the population of WDs created within the GCs as a result of stellar evolution.

We consider two models for the WD population. In the first one, it is assumed that magnetic WDs are created about 11 Gyrs ago as a final products of the evolution of the A and B type stars. We estimate the number of WDs created in the GC by integrating the initial mass function of stars (see Section~2) in the mass range 2-8$M_\odot$. It is expected that during the first 1 Gyrs about
$1.8\times 10^4$ WDs have appeared within the GC with the typical mass of $3\times 10^5$ M$_\odot$.

In the second model, we assume that WDs appear as a result of stellar evolution of stars with the masses in the range 0.8-8 M$_\odot$ with some specific distribution in time. This time  distribution of the WDs is estimated in the following way. 
We apply the initial mass function of stars formed in the GC as derived 
by Salpeter (1955) and Kroupa (2001), i.e. $dN/dM\propto M^{-2.35}$.
On the other hand, we approximate the dependence of the evolution time of the stars with the mass $M$ by $\tau\propto M^{-2.5}$. Based on these 
formulae, we estimate the formation rate of the WDs within the GC as a function of the time, $\tau$, (measured from the formation of the GC),
\begin{eqnarray}
{{dN}\over{d\tau}}\propto {{dN}\over{dM}}\times {{dM}\over{d\tau}}
= {{0.54}\over{\tau_{\rm GC}^{0.54}}}\tau^{-0.46}~~~{\rm Gyrs^{-1}},
\label{eq22}
\end{eqnarray}
\noindent
where the lifetime of the GC is assumed equal to $\tau_{\rm GC} = 12$ Gyrs, and the formation rate of the WDs is normalized to a single WD appearring during the whole lifetime of the GC. Note that the rate of the WD formation within the GC decreases slightly with the age of the GC because the evolution time of less massive stars increases more rapidly with the mass of the star than the factor determining the initial mass function of stars.

\begin{figure*}
\vskip 3.7truecm
\includegraphics{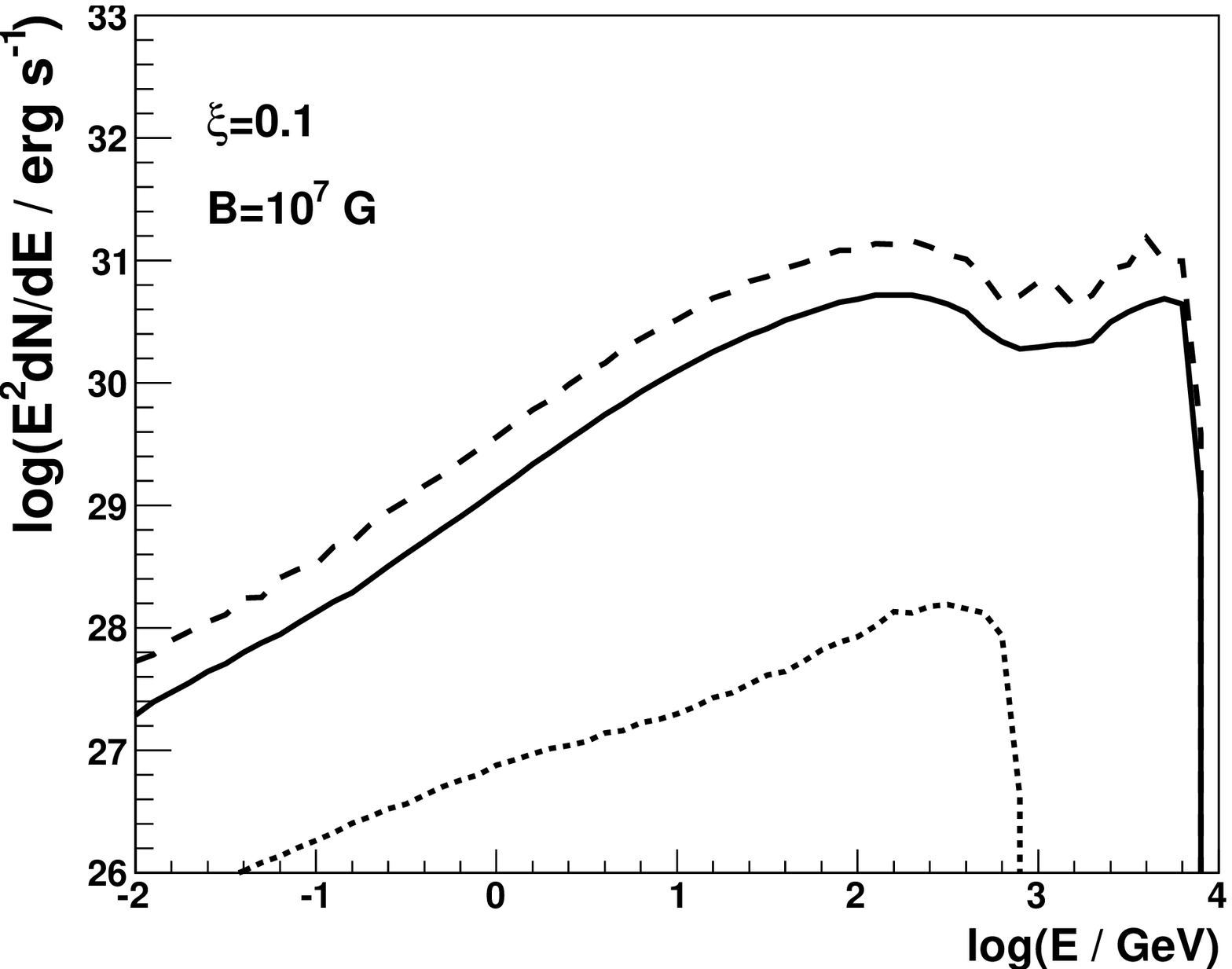}
\includegraphics{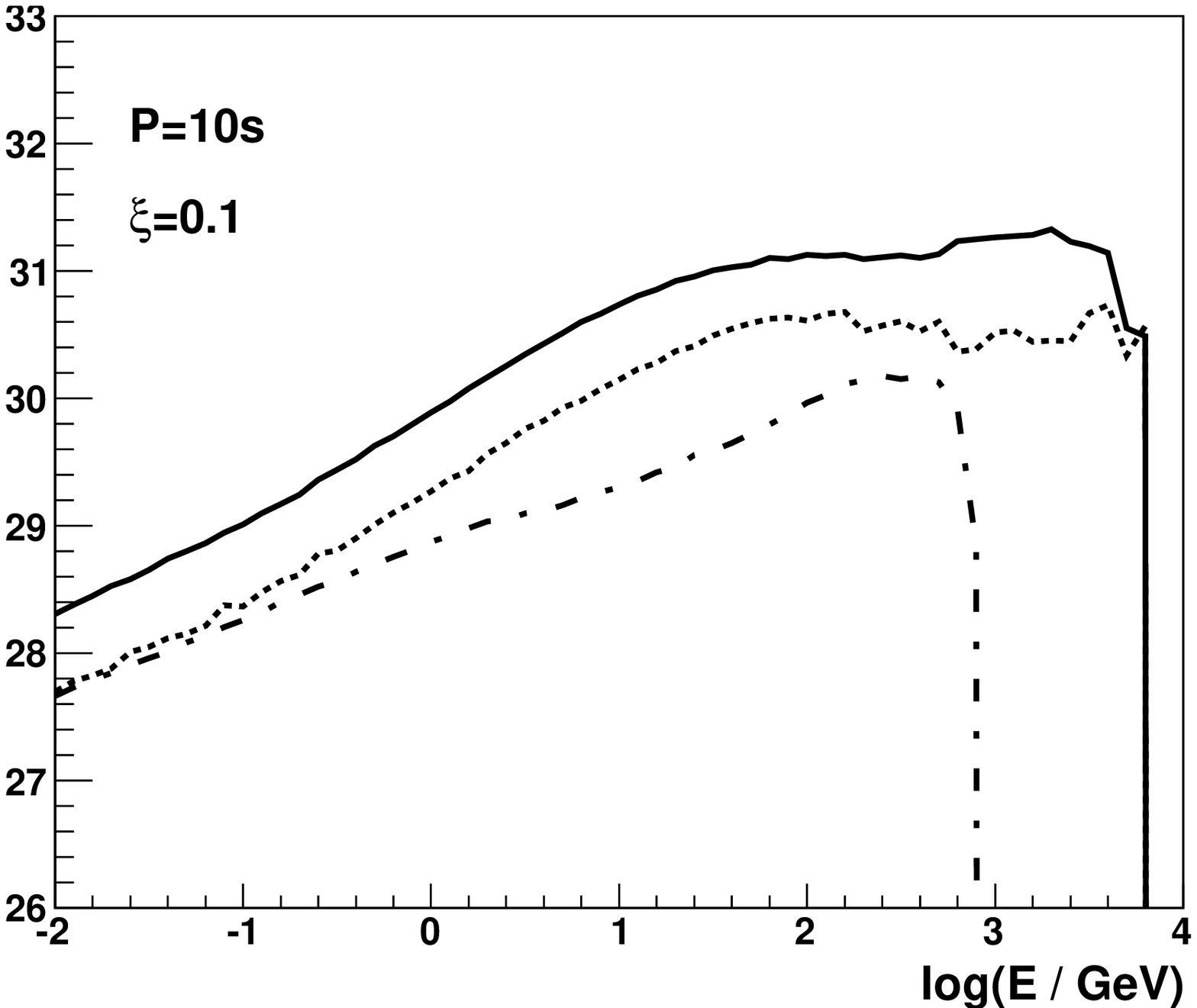}
\includegraphics{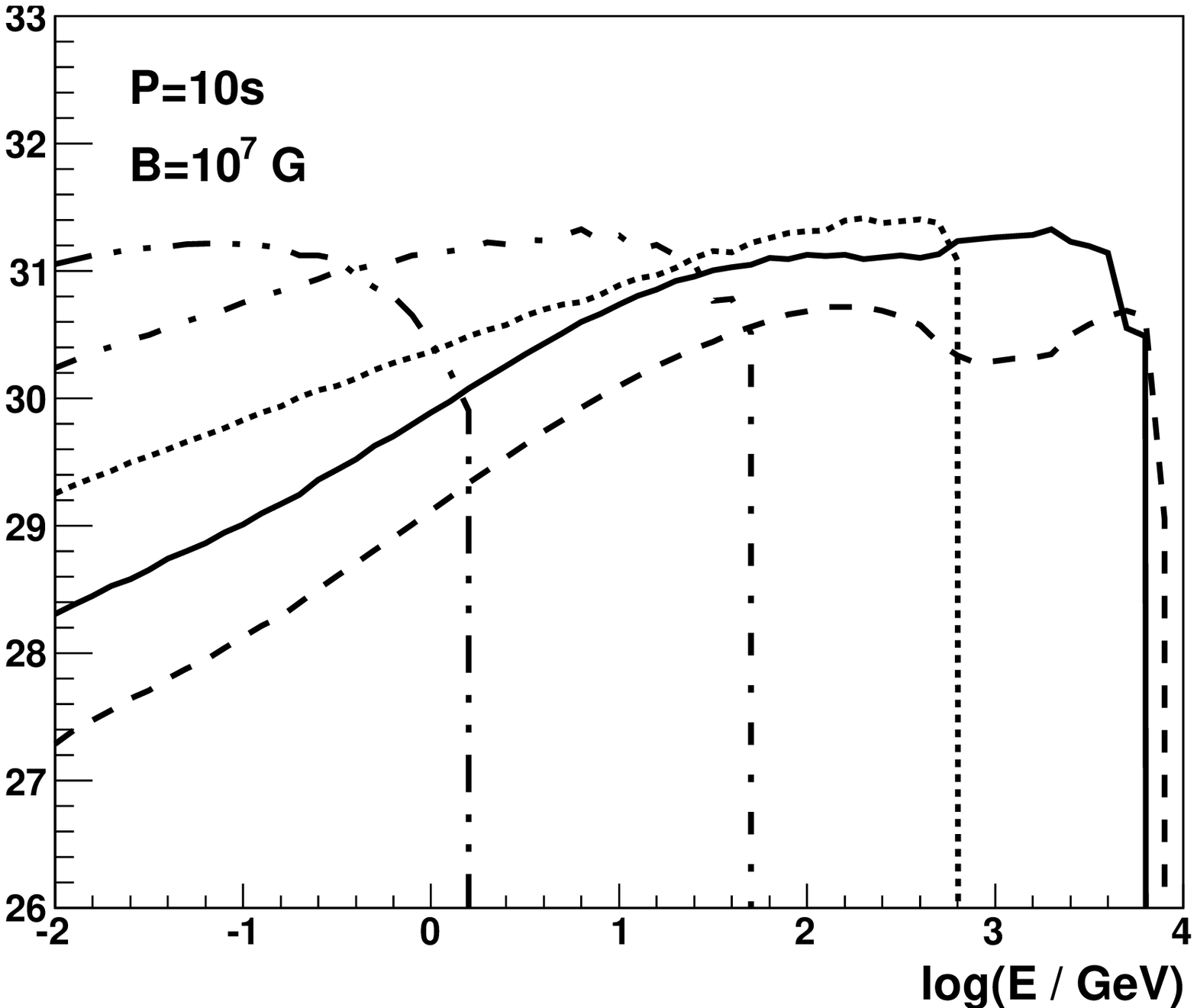}
\caption{As in Fig.~3 but for the WD created during the whole lifetime of the GC equal to 12 Gyrs with the rate given by Eq.~18. Left figure: $\gamma$-ray spectra are shown for the initial period of WDs, $P = 1$ s (dashed) , $10$ s (solid), $30$ s (dot-dashed), and $100$ s (dotted),  their surface magnetic field strength $B = 10^7$ G  and the acceleration efficiency of electrons $\xi = 0.1$. Middle figure: $\gamma$-ray spectra calculated for different strength of the surface magnetic field of the WDs: $10^6$ G (dot-dashed), $10^7$ G (solid), and $10^8$ G (dotted), and for $P = 10$ s, and $\xi = 0.1$. Right figures: Spectra for different acceleration efficiencies: $\xi = 1$ (dashed), $0.1$ (solid), $0.01$ (dotted), $10^{-3}$ (dot-dashed), and $10^{-4}$ (dot-dot-dashed), and $P = 10$ s and $B = 10^7$ G.}
\label{fig4}
\end{figure*}

We calculate the IC $\gamma$-ray spectra produced by relativistic electrons injected from the WD population within the GC for the above defined two models of formation of the WDs. 
The evolution in time of the initial parameters of the WD population is followed as considered in Sect.~2. In the case of the first model, we investigate the dependence of the $\gamma$-ray spectra as a function of the parameters of the WDs
at birth such as the rotational period, the surface magnetic field strength, and the acceleration parameter (see Fig.~3). Depending on these parameters, the $\gamma$-ray spectra usually extend through the GeV-TeV energy range. They are peaked at higher energy end (at TeV energies) due to the domination of the electron energy losses on the IC scattering of the stellar radiation in the Klein-Nishina regime over the scattering of the MBR in the Thomson regime. The $\gamma$-ray spectra only weakly depend on their initial periods,
for the short initial periods of the WDs (in the range of initial periods 1-10 s), due to the saturation of the acceleration process of electrons in the inner WD magnetosphere caused by the curvature energy losses of electrons (see left Fig.~3).
We find interesting dependence of the $\gamma$-ray spectra on the initial strength of the surface magnetic field of the WDs. Note that the spectra reach the highest levels and extend to largest energies for the intermediate values of the magnetic field $B = 10^7$ G. This is due to the dependence of the evolutionary tracks of the WDs with different magnetic field during the time period from their formation up to observation moment (equal to 11 Gyrs). The WDs with weaker magnetic field ($10^6$ G) accelerate electrons with lower efficiency and to lower energies. On the other hand, WDs with stronger magnetic fields ($10^8$ G) are observed at present with longer periods due to more efficient process of rotational energy loss during the 
period from the birth up to the present time.
The shape of the $\gamma$-ray spectra depend strongly on the acceleration coefficient, $\xi$, which determines the energies of electrons injected by the WDs (see right Fig.~3). For the reasonable values of $\xi = 0.1-0.01$,
the $\gamma$-rays from scattering of the stellar and MBR are expected to show the maximum between the GeV  and TeV energies. Note that, electrons scatter mainly stellar radiation in the Thomson regime for $\xi = 0.01$ or in the Klein-Nishina regime for $\xi = 0.1$. In the case of very efficient acceleration of electrons ($\xi = 1$), the contribution to the $\gamma$-ray spectrum from scattering of the stellar radiation in the Klein-Nishina regime and the MBR in the Thomson regime becomes comparable (dashed curve in the right figure).

In the case of the second model, the WDs appear within the GC quite uniformly in time (see Eq.~18). Therefore, WDs are observed at present with parameters more appropriate for efficient acceleration of electrons than in the case of the first model.
WDs produced during the lifetime of the GC appears at present with different rotational periods. They produce $\gamma$-ray spectra extending to higher energies and with larger fluxes in respect to the first model (compare left figures on Fig.~3 and 4). Moreover, the contribution from electrons scattering of the MBR in the Thomson limit is more pronounced in the second model. These general features are also evident in the case of  the $\gamma$-ray spectra produced by the WDs with different injection efficiencies. 
However, in the case of formation of the WDs distributed in time, the $\gamma$-ray spectra extend to clearly larger energies for these same values of the acceleration coefficient $\xi$ due to the presence of relatively more energetic WDs with shorter periods and stronger surface magnetic fields at the observation time.
Note also the difference between the dependence of the $\gamma$-ray spectra on the strength of the magnetic field of the WDs in these two models. In the case of distributed in time formation of the WDs within the GC, the objects with stronger surface magnetic field ($B_{\rm WD} = 10^8$ G) contribute at larger rate to the $\gamma$-ray spectrum than the objects with the weaker surface magnetic field ($B = 10^7$ G).

\subsection{White Dwarfs from mergers in binary systems}

The rate of mergers of the WDs in GCs is difficult to estimate due to the large uncertainties concerning the initial parameters of the WD-WD binary systems and their evolution. We apply the simplest possible assumption that the rate of WD-WD mergers has to depend on the total number of the WDs present at a specific time within the GC, i.e.
\begin{eqnarray}
{{dN^{\rm mer}}\over{d\tau}}\propto\int_{0}^{\tau}{{dN}\over{d\tau'}}d\tau' = {{1.54}\over{\tau_{\rm GC}^{1.54}}}\tau^{0.54}~~~{\rm Gyrs^{-1}}
\label{eq23}
\end{eqnarray}
\noindent
where $dN/d\tau'$ is the rate of the formation of WDs within the GC at the time $\tau'$ as a result of stellar evolution of stars (estimated by Eq.~\ref{eq22}). This rate is also normalized to a single WD created in the merger event. Note that the merger rate of the WDs defined in such a way increases in time starting from the moment of the formation of the GC.
For the WDs formed in the merger events, we apply a more extreme value for the initial surface magnetic field since these WDs have typically much smaller radii (of the order of $\sim 10^8$ cm) which result in a much stronger surface magnetic fields ($\sim 10^9$ G). The initial rotational periods of these WDs are kept in a similar range as in the previous model since such periods are expected from the conservation of the angular momentum of merging WD-WD binary systems (see Sect.~2).

\begin{figure*}
\vskip 3.7truecm
\includegraphics{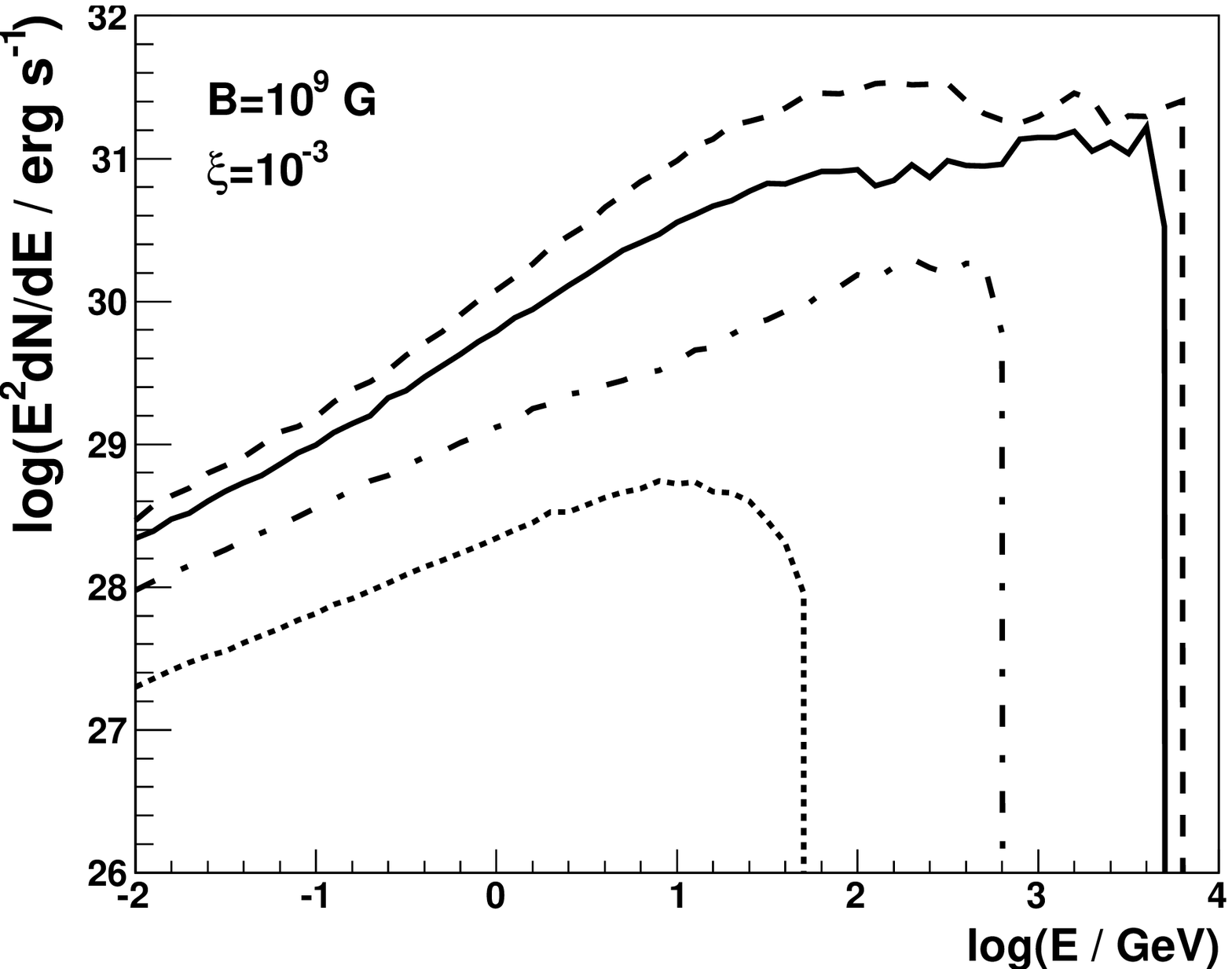}
\includegraphics{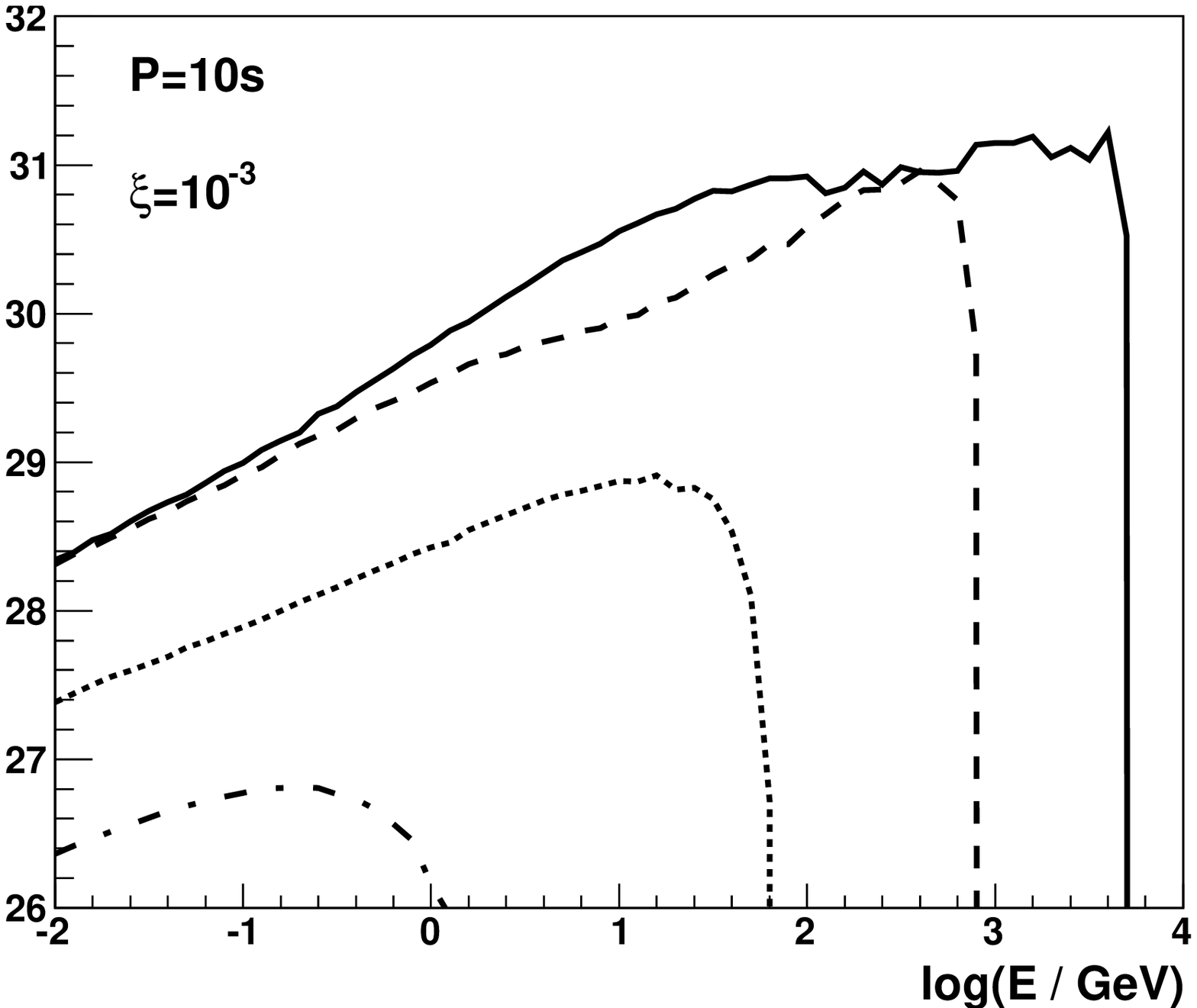}
\includegraphics{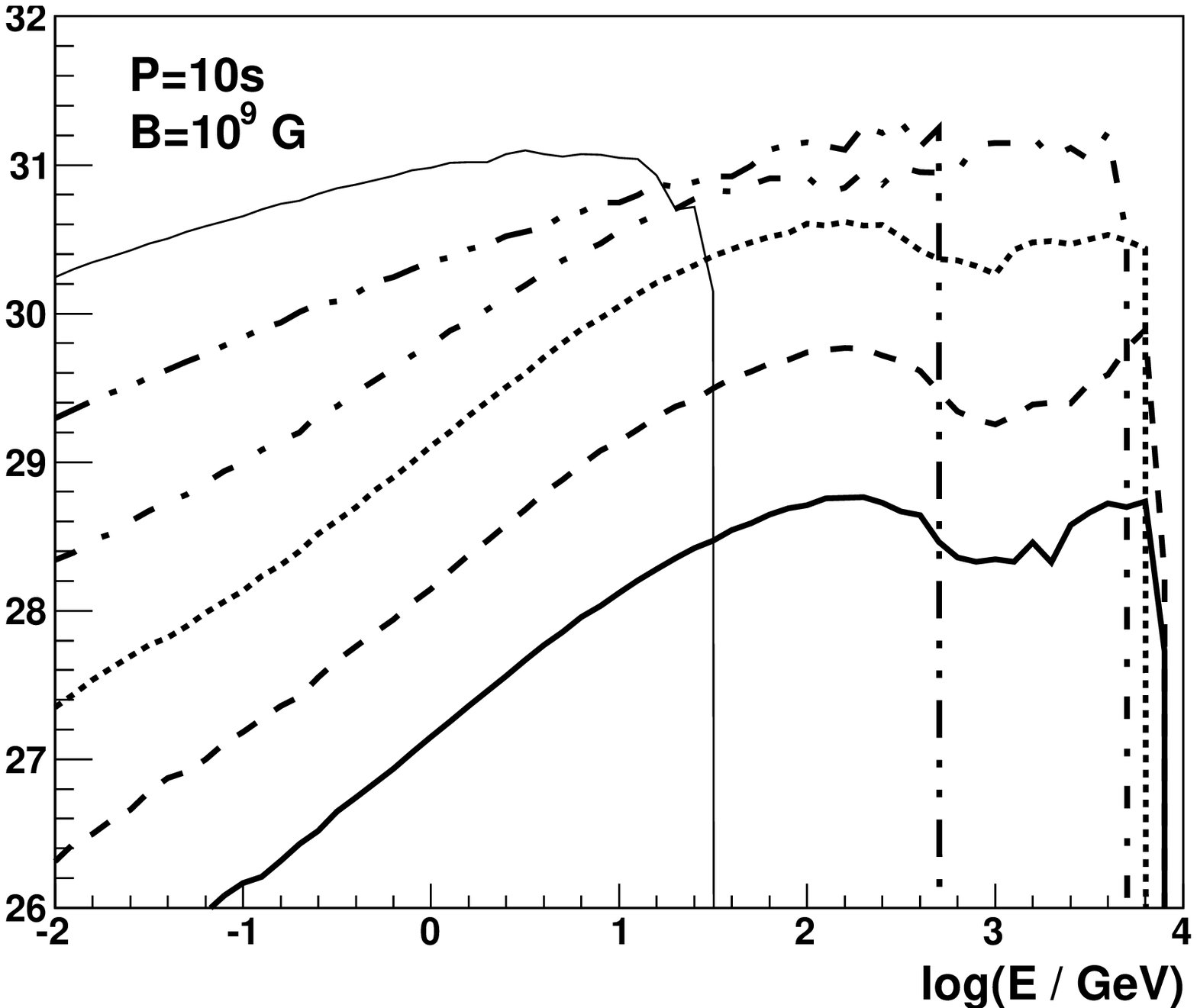}
\caption{As in Fig.~3 but for the WDs created in the merger events of WD-WD binary systems. WDs are injected with the rate given by Eq.~19 during the lifetime of the GC (12 Gyrs). (left figure) $\gamma$-ray spectra are shown for the initial period of the WDs, $P = 1$ s (dashed) , $10$ s (solid), $30$ s (dot-dashed), and $100$ s (dotted), and $B = 10^9$ G and $\xi = 10^{-3}$. 
(middle figure) $\gamma$-ray spectra are calculated for different strength of the surface magnetic field of the WDs: $10^6$ G (dot-dashed), $10^7$ G (dotted), $10^8$ G (dashed), and $10^9$ G (solid), and for $P = 10$ s, and $\xi = 10^{-3}$. (right figure) Spectra for different acceleration efficiencies:
$\xi = 1$ (solid), $0.1$ (dashed), $0.01$ (dotted), $10^{-3}$ (dot-dashed), $10^{-4}$ (dot-dot-dashed), and $10^{-5}$ (thin solid), and $P = 10$ s and $B = 10^9$ G.}
\label{fig5}
\end{figure*}

We calculate the $\gamma$-ray spectra produced by electrons injected from the population of WDs in merger events. We investigate, as in the previous models, the dependence of the $\gamma$-ray spectra on the initial periods of the WDs, the strengths of their surface magnetic field, and the acceleration efficiency (see Fig.~5). The dependence of the $\gamma$-ray spectra on the initial period and the surface magnetic field of the WDs are quite similar to those discussed in the case of the WDs from stellar evolution. However, the $\gamma$-ray fluxes are predicted to be larger due to different dependence of the formation rates of WDs within the GC as a function of time. Significant differences are found in the models with different acceleration efficiencies of electrons in the inner WD magnetospheres. We observe clear saturation of the electron acceleration in a much broader range of acceleration efficiencies (see right Fig.~5 for $\xi$ in the range
$1 - 10^{-3}$). On the other hand, 
the cut-off in the $\gamma$-ray spectra clearly changes only for small values of $\xi$ (see the calculations for $10^{-5}-10^{-4}$).
Due to this saturation effect, the $\gamma$-ray spectra extend only below $\sim 10$ TeV. The saturated acceleration is also responsible for the drastic decrease of the $\gamma$-ray intensity since in these cases significant amount of electron energy is lost already during their propagation in the inner WD magnetosphere.

\subsection{Dependence on the evolution of the magnetic field of the WD}

The calculations of the $\gamma$-ray spectra discussed above are obtained based on the assumption that the surface magnetic field of the WDs is relatively stable on the time scale of the age of the GC. This is not to the end clear as we mentioned in Sect.~2.1. Therefore, we calculate the $\gamma$-ray spectra expected during the diffusion process of electrons within the GC, 
also in the case of a relatively fast decay of the surface magnetic field of the WDs. The decay time scales of the magnetic field in the range, $1-10$ Gyrs, are considered (Fig.~6). Note that the decaying surface magnetic field has not only the effect on the flux and energies of injected electrons by the specific WD but also on the evolution history of the WD in the GC, i.e the evolution of the period of the WD in time also occurs differently (see Eq.~6).
Three basic models for the formation of the WDs within the GC discussed above in detail are considered. We have found strong dependence of the $\gamma$-ray spectra in the case of the model with almost instantaneous formation of the WD population about 11 Gyrs ago (see left Fig.~6). The $\gamma$-ray flux is expected to fall down by about two orders of magnitudes in the case of the WDs in which the magnetic field decays on a time scale of 1 Gyr in comparison to the spectra calculated for the decay time scale of 10 Gyrs. Moreover, the maximum energies of emitted $\gamma$-rays are about an order of magnitude lower. 
The effects caused by the decaying surface magnetic field of the WD on the observed $\gamma$-ray spectra are modest in the case of the formation of WDs as a result of stellar evolution of stars during the whole lifetime of the GC (model II). In this case, we see clear effect of the decaying magnetic field on the $\gamma$-ray flux (decrease by a factor of the order of a few but without significant modification of the shape of the $\gamma$-ray spectrum). 
This is due to a more continuum formation of the WDs during the lifetime of the GC. In the most extreme model, the WDs appear within the GC as a result of mergers in WD-WD binary systems. Such objects are expected to appear mainly latter in the GC when the number density of the WDs in the GC is the largest (see Eq.~19). In such a case, the $\gamma$-ray spectra depend only weakly on the decay time scale of the magnetic field since most of the WDs within GC are relatively young objects. They had no time to significantly change (decay) their surface magnetic field. Therefore, the $\gamma$-ray production is mainly determined by relatively young WDs (with short periods and strong surface magnetic fields) which accelerate electrons to comparable energies (see right Fig.~6).

\begin{figure*}
\vskip 3.7truecm
\includegraphics{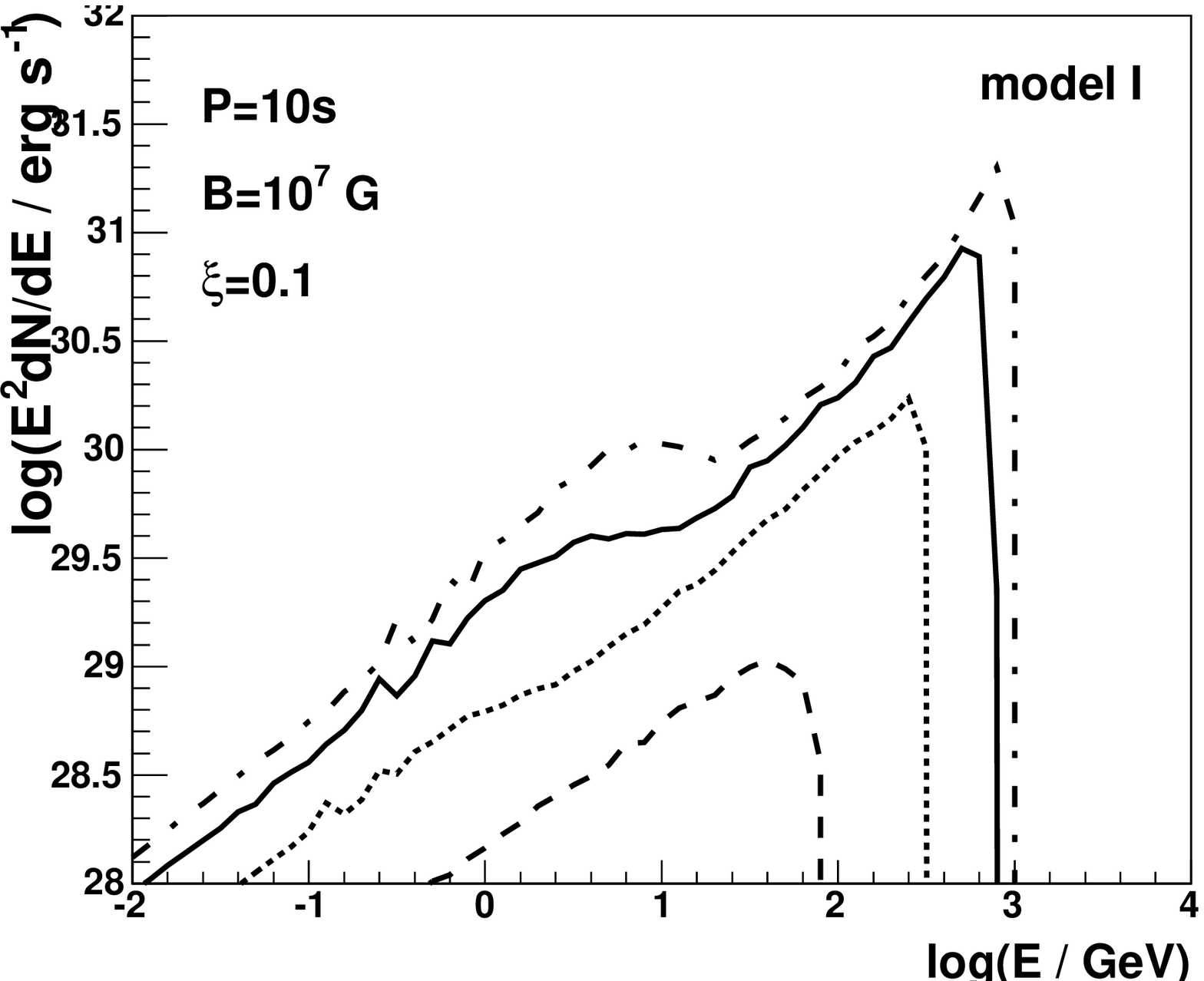}
\includegraphics{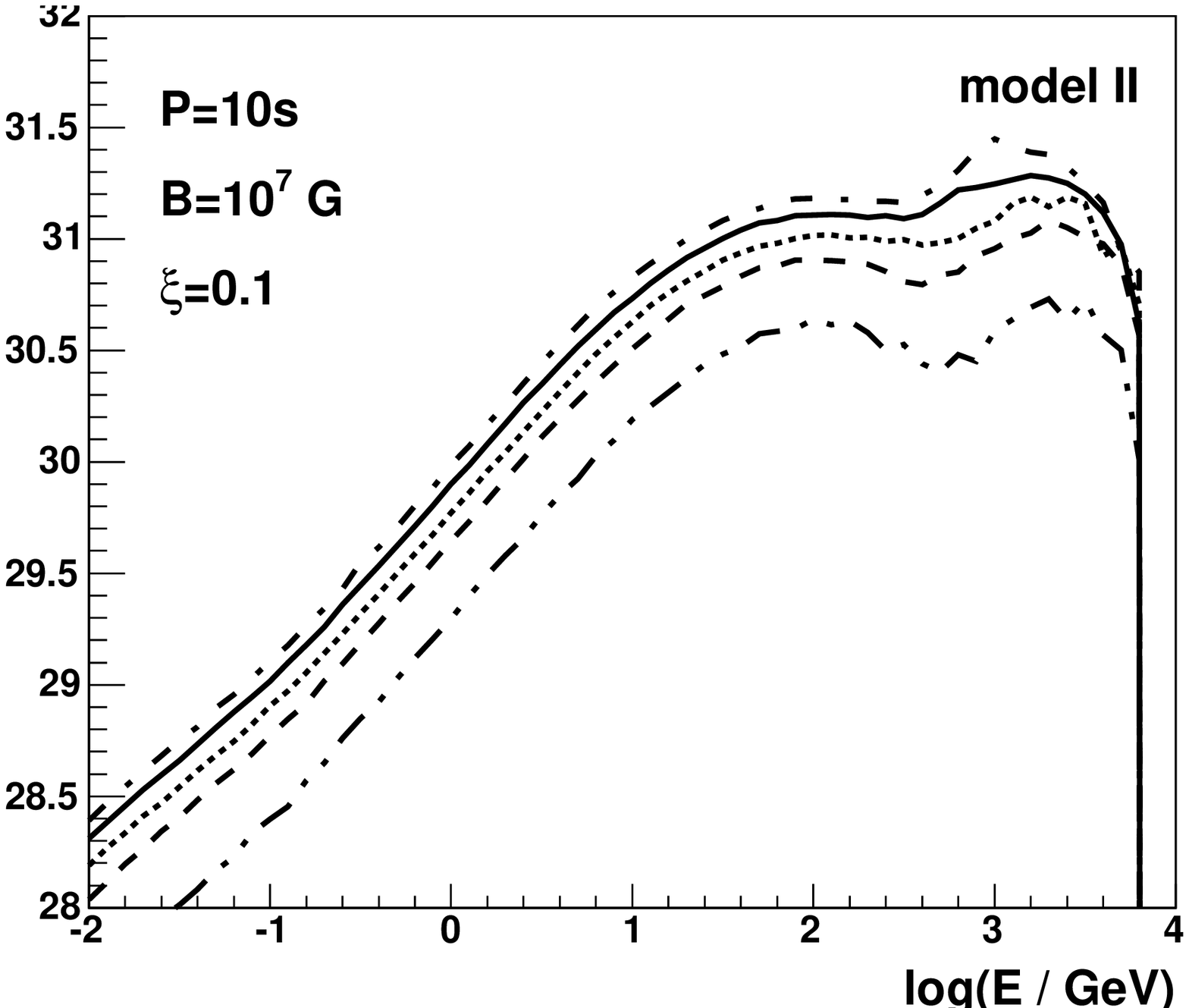}
\includegraphics{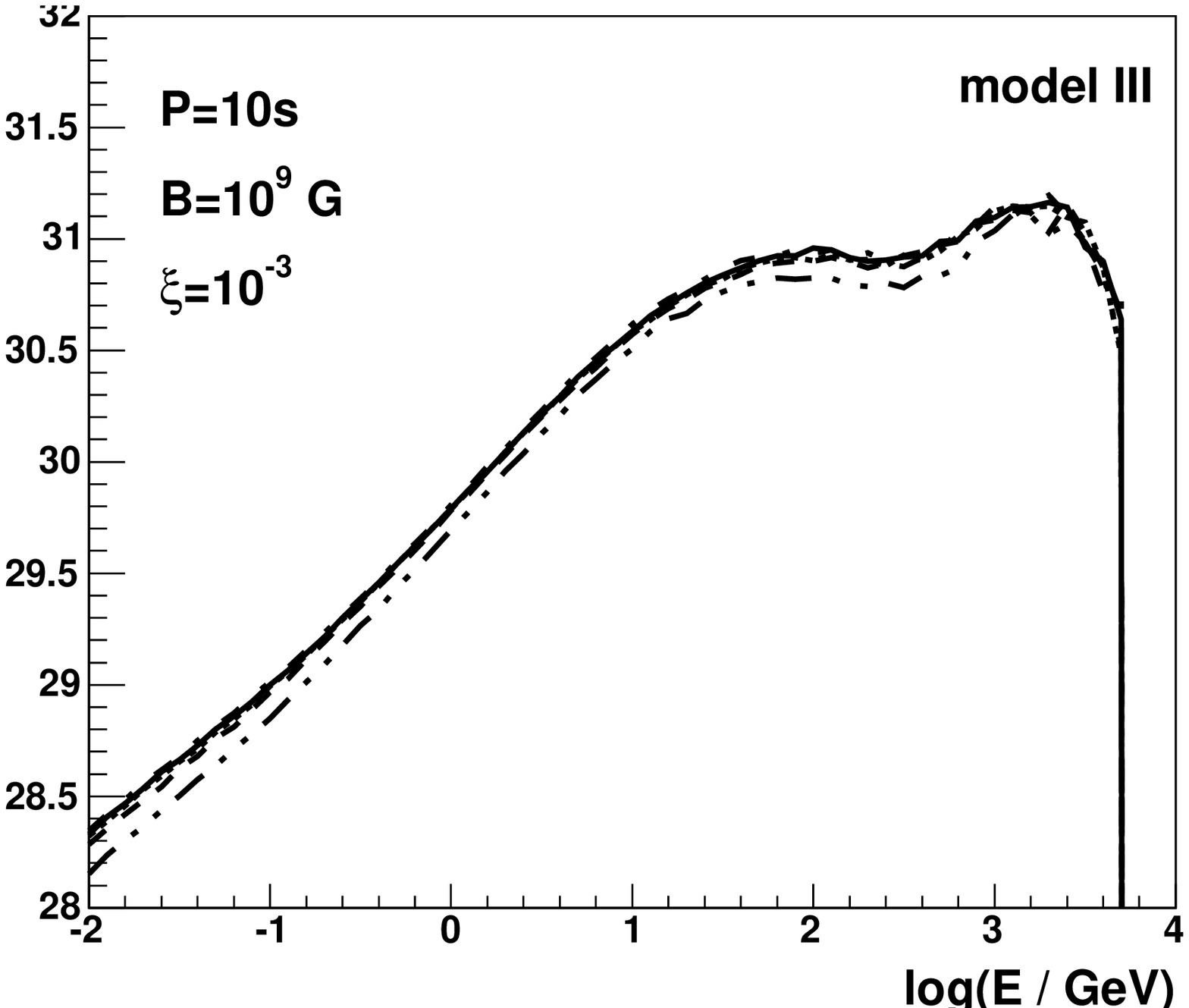}
\caption{Gamma-ray spectra for different decay time of the WD surface magnetic field,
$\tau_{\rm dec} = 1$ Gyrs (dot-dot-dashed curves), $3$ Gyrs (dashed), $5$ Gyrs (dotted), $10$ Gyrs (solid), and without any decay (dot-dashed). Three different models for  the WD formation within the GC are considered:
(left figure) WDs created 11 Gyrs ago with $P = 10$ s, $B = 10^7$ G and $\xi = 0.1$ (model I); (middle) WDs created during the whole lifetime of the GC as a result of stellar evolution with the injection rate given by Eq.~(18) and $P = 10$ s, $B = 10^7$ G  and $\xi = 0.1$ (model II); (right) WDs created in WD-WD mergers during the whole lifetime of GC with the injection rate given by Eq.~(19) and $P = 10$ s, $B = 10^9$ G and $\xi = 10^{-3}$ (model III).}
\label{fig6}
\end{figure*}
\subsection{Dependence on the globular cluster magnetic field}

The magnetic field within the GC determines the diffusion process of injected relativistic electrons (see Sect.~4) and so the rate of their energy losses on the IC process. On the other hand, the magnetic field can also cause additional energy losses of electrons
(synchrotron radiation) which indirectly also influence $\gamma$-ray spectra produced by electrons in the IC process.
We take the synchrotron energy losses of electrons into account when calculating the $\gamma$-ray spectra expected from IC process. The example $\gamma$-ray spectra, obtained in terms of the three considered models for the population of the WDs within the GC, are shown for different strengths of the magnetic field within the GC (see Fig.~7). In the case of the model with instantaneous formation of the WDs 11 Gyrs ago, the differences for the considered range of the magnetic fields ($B_{\rm GC} = 10^{-6}-3\times 10^{-5}$ G) are not very strong. Only in the case with the strongest field the significant amount of energy can be extracted from electrons. 
This relatively small differences are due to the low energies of electrons produced by the old WDs (with the age of 11 Gyrs).
The magnetic field strength has stronger effect on the $\gamma$-ray production in the scattering process of the MBR (first broad bump in the $\gamma$-ray spectrum) since the energy density of this radiation field can become above/below the energy density of the magnetic field in the GC. In the case of the second and third models, the formation of the WDs is distributed during the whole lifetime of the GC. The differences due to the magnetic field within GC are clearly larger in the two last models. 
Note that the $\gamma$-ray spectra with larger fluxes are produced 
for some intermediate values of the magnetic field strength within GC (see $B_{\rm GC}= 3\times 10^{-6}$ G). For weaker fields, the fast diffusion of electrons from the GC results in the decrease of the $\gamma$-ray fluxes. Similar decrease effect is observed in the $\gamma$-ray spectrum calculated for the strong magnetic field within the GC (see $B_{\rm GC} = 3\times 10^{-5}$ G). In this case, the diffusion process is slow but the synchrotron energy losses extract energy from electrons more efficiently. Therefore, we conclude that conditions for the propagation of electrons within the GC considered in the previous figures (for $B_{\rm GC} = 3\times 10^{-6}$ G) are the most optimal for the production of $\gamma$-ray emission in GCs. 
Relatively large dependencies of the $\gamma$-ray spectra on the magnetic field strength within the GC are due to the large energies and extended spectrum of electrons injected by the WD population in terms of the models II and III. In such cases, electrons with different energies diffuse with different speed in the GC. Their synchrotron energy losses also differ significantly.

\begin{figure*}
\vskip 3.7truecm
\includegraphics{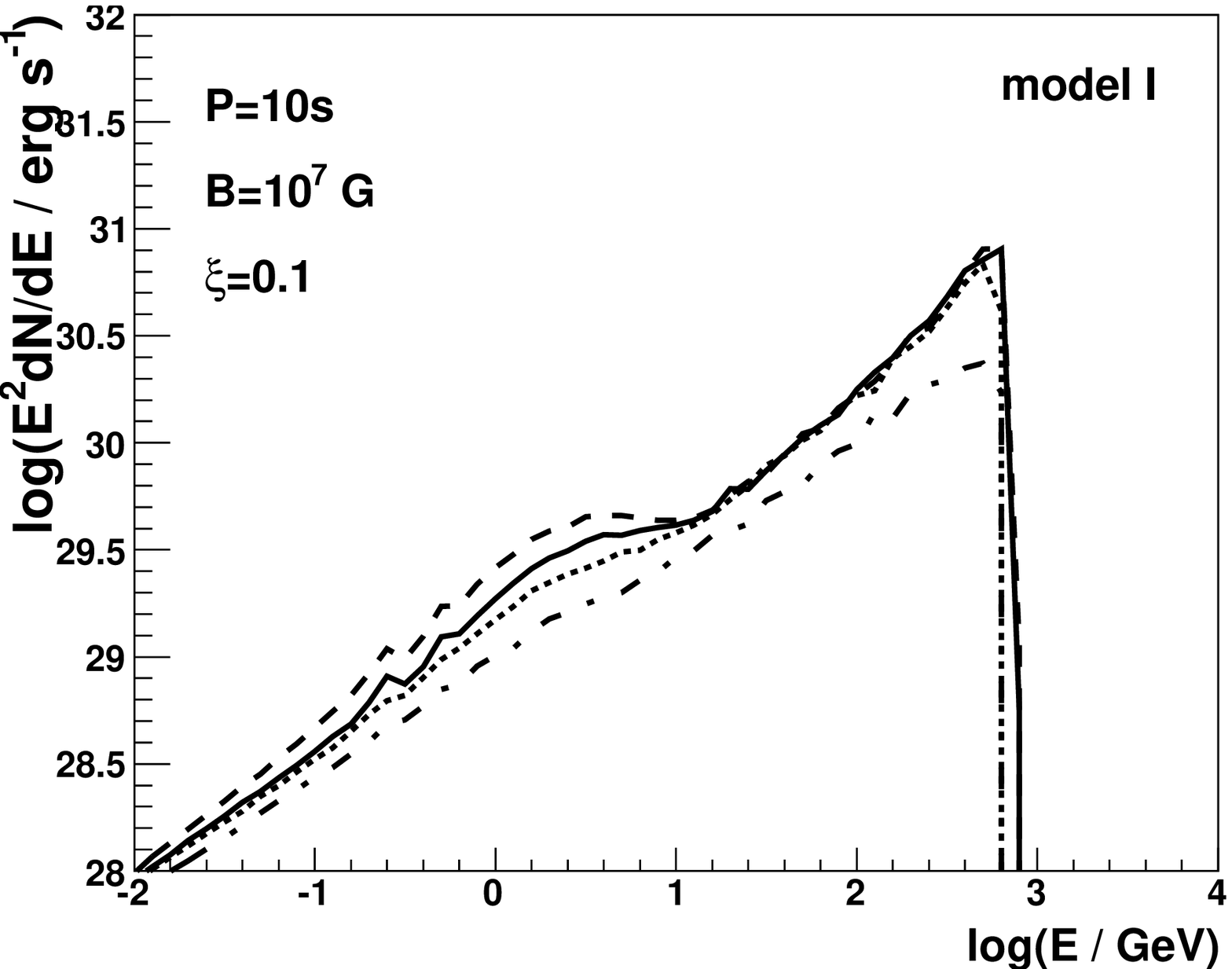}
\includegraphics{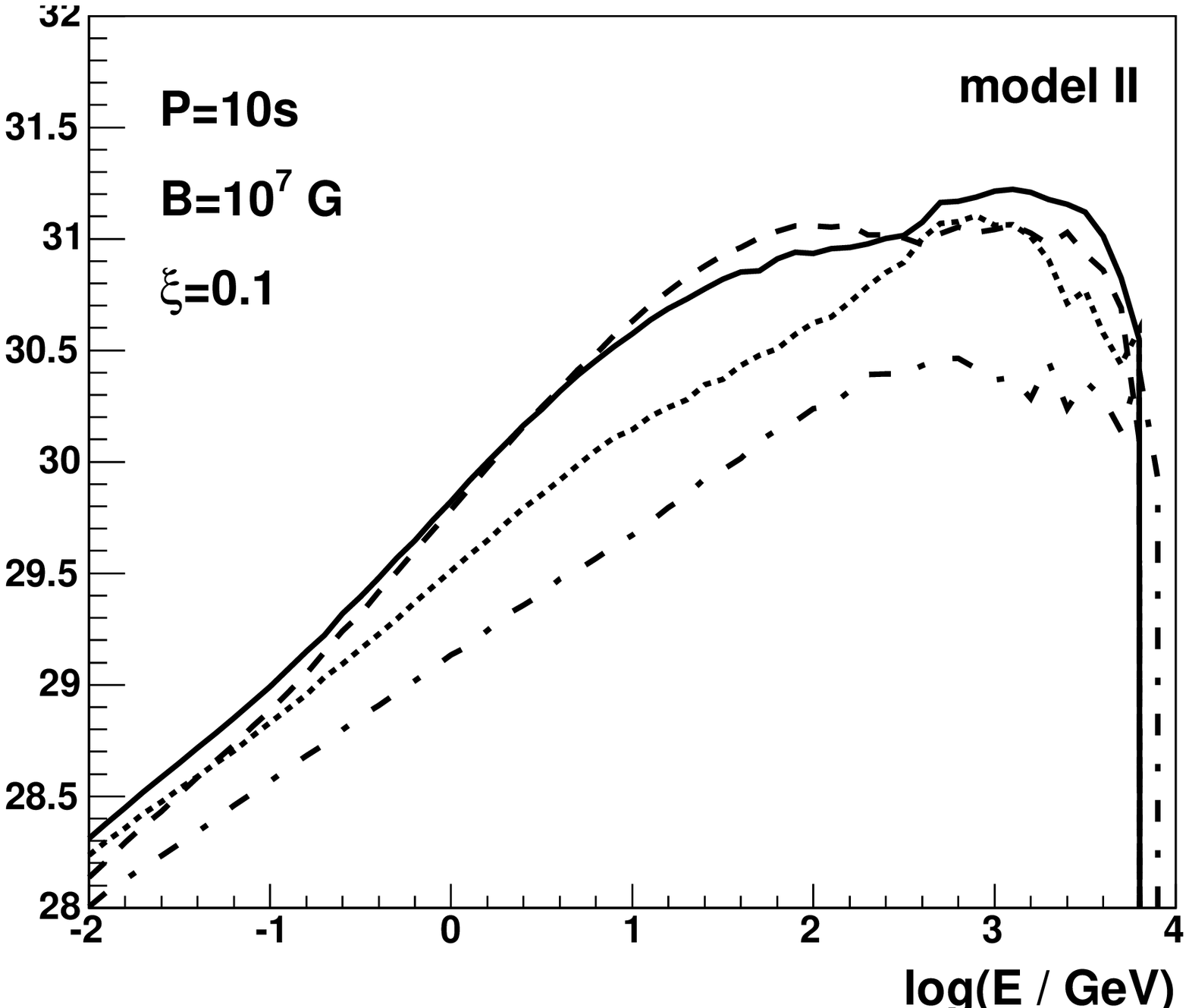}
\includegraphics{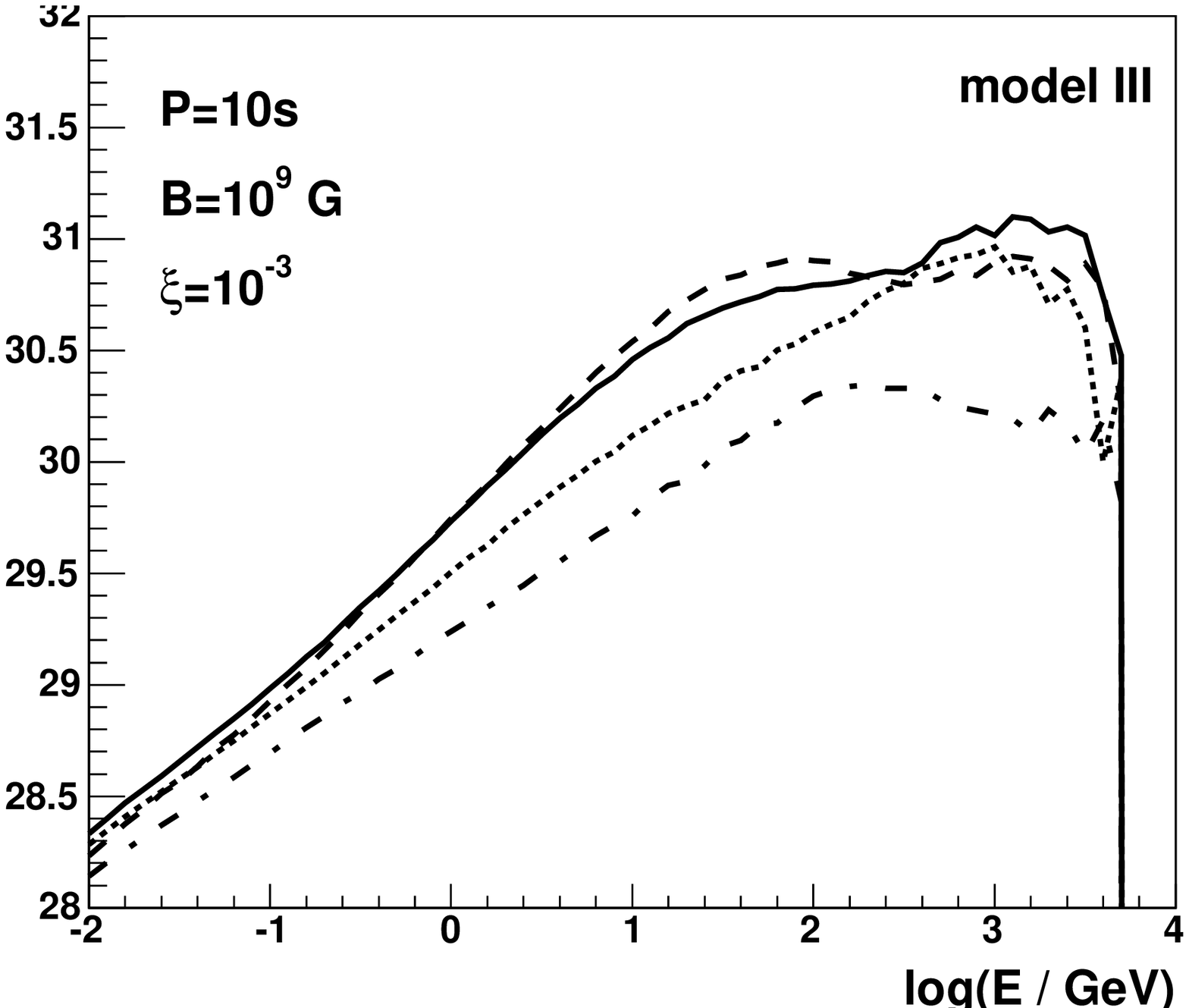}
\caption{Gamma-ray spectra for different magnetic field strengths within the GC,
$B_{\rm GC} = 10^{-6}$ G (dashed curve), $3\times 10^{-6}$ G (solid), $10^{-5}$ G (dotted), and $3\times 10^{-5}$ G (dot-dashed). 
Three different models of the WD creation within the GC are considered:
(left figure) WDs created 11 Gyrs ago with $P = 10$ s, $B = 10^7$ G and $\xi = 0.1$ (model I); (middle) WDs created during the whole lifetime of the GC as a result of stellar evolution with the injection rate given by Eq.~(18) and $P = 10$ s, $B = 10^7$ G  and $\xi = 0.1$ (model II); (right) WD created in WD-WD mergers during the lifetime of GC with the injection rate given by Eq.~(19) and $P = 10$ s, $B = 10^9$ G and $\xi = 10^{-3}$ (model III).
All spectra are normalized to $N_{\rm WD}\times \eta = 1$. The decay time of the surface magnetic field of the WDs is equal to 10 Gyrs.}
\label{fig7}
\end{figure*}
\section{Prospects for detection of $\gamma$-rays with Cherenkov telescopes}

Up to now, we have investigated the $\gamma$-ray fluxes and spectra in the case of electron injection by the supposed single WD within the GC. As we have estimated above the number of the WDs, formed as a final products of evolution of stars can be as large as a few tens of thousands. Therefore, in order to estimate the expected fluxes of the $\gamma$-rays from specific GC we have to re-normalize the spectra in Figs.~3-7 to the realistic product of the number of magnetized WD within the GC and the energy conversion efficiency from rotating WD to relativistic electrons.
In the case of the WD population, we assume that the number of the magnetized WDs is approximately equal to about $10\%$ of the number of all formed WDs within the GC (see such estimates in the proximity of the Sun discussed in Sect.~2). In such a case, the number of magnetized WDs in specific GC is expected to be as large as a few thousand. 
The acceleration processes of electrons in the magnetspheres of rotating WDs should be analogical to those expected in the magnetospheres of rotating neutron stars. Therefore, we assume (following the modelling of pulsars) that a part of the rotational energy transferred to relativistic electrons in the WD magnetospheres can be of the order of $\eta\sim 10\%$. 
Note however, that electrons accelerated in the inner MSP magnetospheres
lose already there the essential part of their energy on emission of pulsed $\gamma$-rays. This pulsed emission from the population of the MSPs within the GCs has been recently detected by the Fermi-LAT detector in the GeV energy range. On the other hand, electrons can escape from the inner WD magnetospheres with the negligible energy losses. In conclusion, It is likely that the realistic values of the product $N_{\rm WD}\times \eta$ can be of the order of $\sim 10^3$.
We assume that the number of WDs from mergers of WD-WD binary systems can be of the order of a few hundred, i.e. $10\%$ of all WDs are in such binary systems and about $10\%$ of them already merged. 
Then, the values of the product $N_{\rm WD}\times \eta$ can be in this case of the order of $\sim 10^2$. 

The $\gamma$-ray spectra calculated in the considered model are typically expected on the level of $\sim 10^{31}$ erg s$^{-1}$, assuming that the product, $N_{\rm WD}\times \eta$, of the number of WDs times the energy conversion efficiency from WDs to electrons equals unity (see Figs.~3-7). Based on the available
upper limits and possible detections of GCs by the operating Cherenkov telescopes, we can try to 
constrain this free parameter of the model. As an example, we use the upper limits reported by the HESS Collaboration on 47 Tuc (Aharonian et al.~2009) equal to $6.7\times 10^{-13}$ erg cm$^{-2}$ s$^{-1}$ above 800 GeV, and by the MAGIC Collaboration on M 13 (Anderhub et al.~2009) equal to $5.1\times 10^{-12}$ cm$^{-2}$ s$^{-1}$ above 200 GeV (assuming the differential $\gamma$-ray spectrum with the spectral index equal to -2.5). We also use the flux reported for the TeV source close to Ter 5 by the HESS Collaboration (Abramowski et al.~2011b) equal to $1.2\times 10^{-12}$ cm$^{-2}$ s$^{-1}$ above 440 GeV 
The comparison of the observations with the calculated $\gamma$-ray spectra allows us to constrain the product $N_{\rm WD}\times \eta$ to be below $\sim 600$ for 47 Tuc and $\sim 1500$ for M 13. We also estimate the value of this product on $\sim 900$ for Ter 5 by assuming that this emission really comes from this GC. These values start to constrain the values of $N_{\rm WD}\times \eta$ estimated above.

Let us now compare the expected $\gamma$-ray fluxes from the GC at a typical distance scale with the expected sensitivity of the future telescopes. 
The Cherenkov Telescope Array (CTA) is expected to have 50 hr differential sensitivity of the order of $10^{-13}$ erg cm$^{-2}$ s$^{-1}$ at the TeV energies (configuration E, see Actis et al.~2011).
Therefore, it is expected that the source at a typical distance of GC (10 kpc) can be detected if its differential $\gamma$-ray spectrum is on the level of $E^2dN/dE = 10^{33}$ erg cm$^{-2}$ s$^{-1}$.
The $\gamma$-ray spectra shown in Figs.~3-7 should be re-normalized by the factors $10^3$ in the case of the WDs from stellar evolution and $10^2$ in the case of WDs from mergers (values estimated above), before their direct comparison with the sensitivity limit of CTA. We conclude that in the case of electron acceleration by the WDs produced about 11 Gyrs ago within the GC (model I), the $\gamma$-ray emission might be detected provided that WDs are formed with the initial periods not significantly longer than $\sim 10$ s, the initial surface magnetic field is above $\sim 10^6$ G (but not much stronger than $10^7$ G), the decay time scale of the magnetic field is larger than $\sim 5$ Gyrs, and the acceleration efficiency of electrons in the WD inner magnetosphere is close to $\sim 0.1$ or larger (see Figs.~3 and 6). In the case of the model with continuous formation of WDs within the GC as a result of the stellar evolution (model II),
the conditions for detection of the $\gamma$-ray emission are clearly less restrictive since many WDs appear relatively late after the GC formation.  In this case the $\gamma$-rays produced by WDs with initial periods of the order of $100$ s might have chance to be detected by CTA (see Fig.~4). WDs with the periods above 10 s could produce detectable fluxes of $\gamma$-rays 
for the whole investigated range of the surface magnetic fields at birth
(i.e. $10^6-10^8$ G). For such WDs, electrons accelerated in the models described by the acceleration parameters above $\sim 10^{-3}$ should produce $\gamma$-rays at energies above the detection threshold of CTA. The $\gamma$-ray fluxes (model II) are above sensitivity of CTA even for a relatively short decay time scale of the surface magnetic field strength of the WDs. This is due to the fact that many WDs are formed at the latter period after the GC formation (see Fig.~6). 
On the other hand, the situation is not so optimistic in the case of the third model in which the WD population appears as a result of mergers of the WD-WD binary systems. The number of such WDs from mergers is about an order of magnitude lower than in the previous models. The $\gamma$-ray spectra are in this case on the sensitivity limit of CTA for WDs with the periods of the order of 10 s (and shorter), the surface magnetic fields in therange $10^8-10^9$ G, and the acceleration efficiencies in the range between a few $10^{-4}$ and $10^{-3}$ (see Fig.~5). Since in this model most of the merger events occur late during the lifetime of the GC, the spectra weakly depend on the decay time scale of the surface magnetic field strength (see right Fig.~6). 

The magnetic field strength within the GC have a minor effect on the level of the $\gamma$-ray fluxes and their spectra in the case of the formation of the WDs instantaneously at 11 Gyrs ago since these electrons have relatively low energies at the present time. However, significant dependence of the $\gamma$-ray spectra is found in the case of the extended in time formation of the WDs within GC in terms of the models II and III (WDs from stellar evolution and from mergers). The largest $\gamma$-ray fluxes in the TeV energy range are obtained in the case of the intermediate values of the magnetic field strengths within the GC (see Fig.~7). For weaker fields, the diffusion process is fast  but the synchrotron energy losses are negligible. For strong fields, the diffusion is slow, keeping relativistic electrons in the dense thermal radiation field, but the synchrotron losses becomes important in comparison to the IC energy losses on scattering of the MBR and stellar radiation. Therefore, electrons lose significant amount of energy on the synchrotron process in this last case.
As a result, the scattering of stellar radiation by electrons becomes less efficient and produced $\gamma$-ray spectra show maximum at lower energies. Moreover, the scattering of the MBR becomes inefficient (energy density of the magnetic field dominates over the energy density of the MBR).

\section*{Acknowledgments.}
\noindent
This work is supported by the grant through the Polish Narodowe 
Centrum Nauki No. 2011/01/B/ST9/00411.

\section*{References}

\noindent
Abdo, A.A. et al. 2009a {\it Science} {\bf 325} 845

\noindent
Abdo, A.A. et al. 2009b {\it Science} {\bf 325} 848 

\noindent
Abdo, A.A. et al. 2010 {\it Astron.Astrophys.}  {\bf 524} 75

\noindent
Abdo, A.A. et al. 2011 {\it Science} DOI: 10.1126/science.1207141  

\noindent
Abramowski, A. et al. 2011a {\it Astrophys.J.} {\bf 735} 12

\noindent
Abramowski, A. et al. 2011b {\it Astron.Astrophys.} {\bf 531} 18  

\noindent
Actis, M. et al. 2011 {\it Exp.Astron.} {\bf 32} 193

\noindent
Aharonian , F.A. et al. 2009 {\it Astron.Astrophys.} {\bf 499} 273  

\noindent
Anderhub et al. 2009 {\it Astrophys.J.} {\bf 702} 266

\noindent
Arons, J. , Scharlemann, E.T. 1979 {\it Astrophys. J.} {\bf 231} 854

\noindent
Bednarek, W. 2011 Proc. {\it First Session of the Sant Cugat Forum on Astrophysics}, eds. N. Rea \& D.F. Torres (Springer) p. 185

\noindent
Bednarek, W., Sitarek, J. 2007 {\it Mon.Not.R.Astr.Soc.} {\bf 377}  920

\noindent
Blumenthal, G.R., Gould, R.J. 1970 {\it Rev.Mod.Phys.} {\bf 42} 237

\noindent
Cheng, K.S., Chernyshov, D.O., Dogiel, V.A., Hui, C.Y., Kong, A.K.H. 2010 {\it Astrophys.J.} {\bf 723} 1219

\noindent
De Jager, O.C. 1994 {\it Astrophys.J.Suppl.} {\bf 90} 775

\noindent
Goldreich, P., Julian, H. 1969 {\it Astrophys.J.} {\bf 157} 869

\noindent
Grindlay, J. E., Heinke, C.O., Edmonds, P.D., Murray, S.S., Cool, A.M. 2001 {\it Astrophys.J.} {\bf 563} L53

\noindent
Harding, A.K., Muslimov, A. 1998 {\it Astrophys.J.} {\bf 508} 328

\noindent
Harding, A.K., Usov, V.V., Muslimov, A. 2005 {\it Astrophys.J.} {\bf 622} 531

\noindent
Harris, W.E. 1996 {\it Astron.J.} {\bf 112} 1487

\noindent
Heinke, C.O. 2011, Proc. {\it Binary Star Evolution: Mass Loss, Accretion and Mergers}, eds. V. Kalogera and M. van der Sluys (AIP Conf. Ser., Mykonos, Greece) (arXiv:1101.5356)

\noindent
Hui, C. Y., Cheng, K. S., Taam, R. E. 2010 {\it Astrophys.J.} {\bf 714} 1149

\noindent
Ikhasanov, N.R. 1998 {\it Astron.Astrophys.} {\bf 338} 521

\noindent
Ivanova, N., Belczy\'nski, K., Fregeau, J.M., Rasio, F.A. 2005 {\it Mon.Not.R.Astr.Soc.} {\bf 358} 572

\noindent
Kong, A. K. H., Hui, C. Y., Cheng, K. S. 2010 {\it Astrophys.J.} {\bf 712} 36 

\noindent
Kroupa, P. 2001 {\it Mon.Not.R.Astr.Soc.} {\bf 322} 231

\noindent
Muslinov, A.G., Van Horn, H.M., Wood, M.A. 1995 {\it Astrophys.J.} {\bf 442} 758

\noindent
Paczy\'nski, B. 1990 {\it Astrophys.J.} {\bf 365} L9

\noindent
Pellizzoni, A. et al. 2009 {\it Astrophys.J.} {\bf 695} L115

\noindent
Ruderman, M.A., Sutherland, P.G. 1975 {\it Astrophys.J.} {\bf 196} 51

\noindent
Salpeter, E.E. 1955 {\it Astrophys.J.} {\bf 121} 161

\noindent
Sefako, R.R., de Jager, O.C., 2003 {\it Astrophys.J.} {\bf 593} 1013 

\noindent
Usov, V.V. 1993 {\it Astrophys.J.} {\bf 410} 761

\noindent
Venter, C., de Jager O.C. 2008  {\it Astrophys.J.} {\bf 680} L125 

\noindent
Venter, C., de Jager O.C. \& Clapson, A.C. 2009 {\it Astrophys.J.} {\bf 696} L52 

\noindent
Wendell, C.E., Van Horn, H.M., Sargent, D. 1987 {\it Astrophys.J.} {\bf 313} 284

\noindent
Wickramasinghe, D.T., Ferrario, L. 2000 {\it Publ.Astr.Soc.Pac.} {\bf 112} 873

\end{document}